\documentclass[review,sort&compress]{elsarticle}

\usepackage{lineno,hyperref,fullpage}

\journal{Computers \& Fluids}

\usepackage{nomencl}
\usepackage{amsmath}
\usepackage{amssymb}
\usepackage{multirow}
\usepackage{subcaption}

\usepackage{floatrow}
\makenomenclature
\usepackage{xpatch}
\xpatchcmd{\thenomenclature}{\section*{\nomname}
}{}{\typeout{Success}}{\typeout{Failure}}
\RequirePackage{ifthen}
\printnomenclature[2cm]

\usepackage[utf8]{inputenc}
\usepackage{soul}
\usepackage{xcolor}
\usepackage{bm}
\usepackage{natbib}

\usepackage{etoolbox}

\begin{document}

\begin{frontmatter}

\title{Evaluation of ensemble methods for quantifying uncertainties in steady-state CFD applications with small ensemble sizes}

\author[lmfl,vt]{Xinlei Zhang}

\author[vt]{Heng Xiao\corref{mycor}}
\cortext[mycor]{Corresponding author}
\ead{hengxiao@vt.edu}
\ead[url]{https://www.aoe.vt.edu/people/faculty/xiaoheng.html}

\author[lmfl]{Thomas Gomez}

\author[lmfl,vt]{Olivier Coutier-Delgosha}

\address[lmfl]{Univ. Lille, CNRS, ONERA, Arts et Metiers Institute of Technology, Centrale Lille, UMR 9014 - LMFL - Laboratoire de Mécanique des fluides de Lille - Kampé de Fériet, F-59000 Lille, France}
\address[vt]{Kevin T. Crofton Department of Aerospace and Ocean Engineering, Virginia Tech, Blacksburg, VA 24060, USA}

\begin{abstract}
Bayesian uncertainty quantification (UQ) is of interest to industry and academia as it provides a framework for quantifying and reducing the uncertainty in computational models by incorporating available data.
For systems with very high computational costs, for instance, the computational fluid dynamics (CFD) problem, the conventional, exact Bayesian approach such as Markov chain Monte Carlo is intractable. 
To this end, the ensemble-based Bayesian methods have been used for CFD applications.
However, their applicability for UQ has not been fully analyzed and understood thus far.
Here, we evaluate the performance of three widely used iterative ensemble-based data assimilation methods, namely ensemble Kalman filter, ensemble randomized maximum likelihood method, and ensemble Kalman filter with multiple data assimilation for UQ problems.
We present the derivations of the three ensemble methods from an optimization viewpoint.
Further, a scalar case is used to demonstrate the performance of the three different approaches with emphasis on the effects of small ensemble sizes.
Finally, we assess the three ensemble methods~for quantifying uncertainties in steady-state CFD problems involving turbulent mean flows. Specifically, the Reynolds averaged Navier--Stokes (RANS) equation is considered the forward model, and the uncertainties in the propagated velocity are quantified and reduced by incorporating observation data.
The results show that the ensemble methods cannot accurately capture the true posterior distribution, but they can provide a good estimation of the uncertainties even when very limited ensemble sizes are used.
Based on the overall performance and efficiency from the comparison, the ensemble randomized maximum likelihood method is identified as the best choice of approximate Bayesian UQ approach~among the three ensemble methods evaluated here.
\end{abstract}

\begin{keyword}
Uncertainty quantification \sep Ensemble methods \sep Data assimilation \sep Computational fluid dynamics \sep Small ensemble sizes
\end{keyword}

\end{frontmatter}

\section{Introduction}
\subsection{Bayesian uncertainty quantification for CFD}
\begin{figure}[!htbp]
    \centering
    \includegraphics[width=0.8\textwidth]{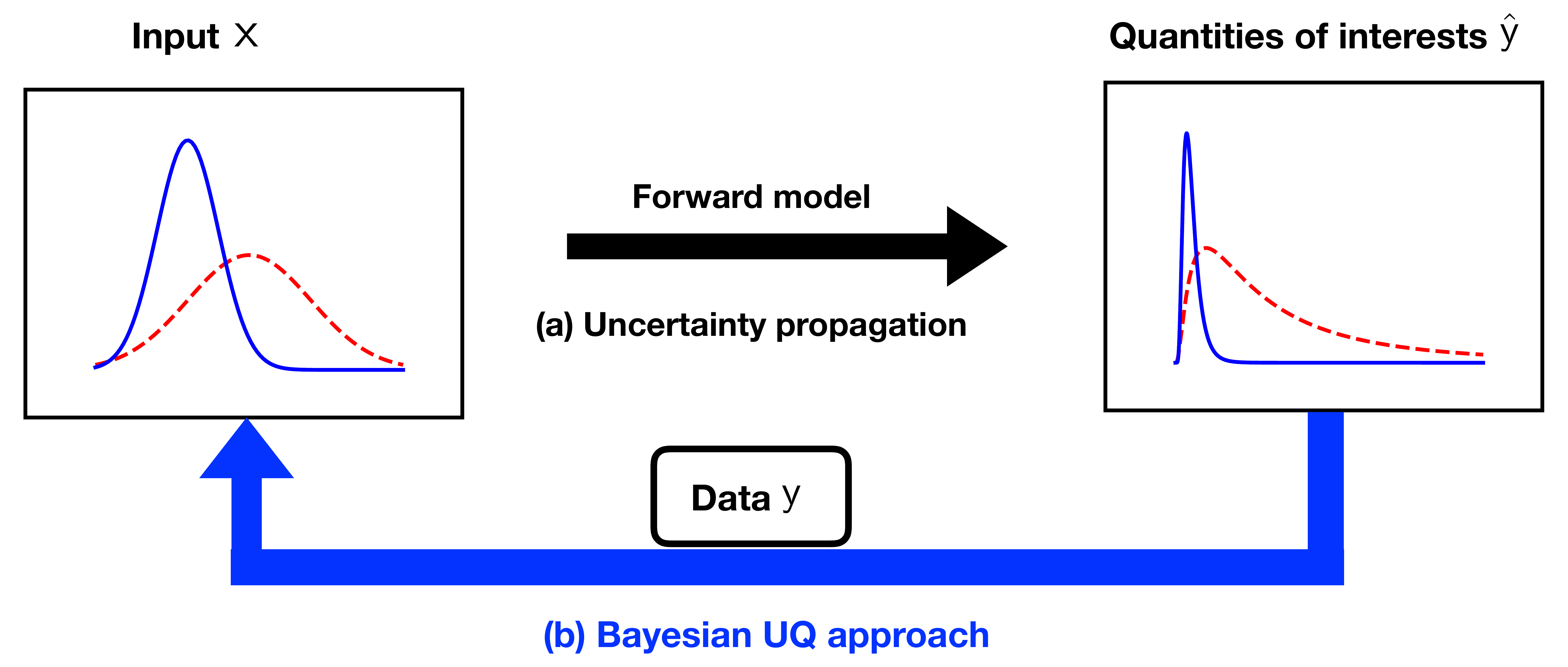}
    \caption{Schematic of uncertainty quantification. (a) Uncertainty propagation forwardly propagates the presumed uncertainty (red dashed line) in the input to the quantity of interest through the forward model; (b) Bayesian UQ approach combines the prior information with high fidelity simulation or experimental data to backwardly quantify the posterior uncertainty (blue solid line) in the quantities of interest and in the input.}
    \label{fig:Bayes_UQ}
\end{figure}

In computational fluid dynamics (CFD) applications, Reynolds averaged Navier--Stokes (RANS) methods still are the workhorse tool to inform the important decision-making during engineering design processes. However, RANS models cannot provide accurate results for many cases in the presence of complex turbulent flows. That necessitates quantifying uncertainties in the numerical simulations so that we could obtain additional confidence/statistics information on the simulated results~\cite{witherden2017future}. 
The conventional approach to quantify uncertainties is to forwardly propagate the presumed uncertainty in system inputs to the quantity of interests (QoIs) through the forward model.
The procedure of the uncertainty propagation is illustrated in Fig.~\ref{fig:Bayes_UQ}(a). Numerous methods~\cite{ mathelin2004uncertainty, knio2006uncertainty, hosder2006non} and applications~\cite{najm2009uncertainty, dow2011quantification, gel2013applying} have been developed and explored for uncertainty propagation in the literature.
Another uncertainty quantification (UQ) method is Bayesian UQ approach. This approach can account for the available data from high fidelity simulations or experiments to backwardly quantify and reduce the uncertainty of QoIs as well as the system inputs (e.g., model parameters or underlying terms)~\cite{xiao_quantifying_2016}. The procedure of Bayesian UQ is illustrated in the schematic in Fig.~\ref{fig:Bayes_UQ}(b).

Numerous works have been conducted to apply Bayesian UQ approach to diverse applications, including RANS simulations.
Based on the pioneering work of Kennedy and O'Hagan~\cite{kennedy2001bayesian},
Cheung et al.~\cite{cheung2011bayesian} applied a Bayesian calibration framework for the Spalart--Allmaras turbulence model to calibrate the model parameters by incorporating experimental measurements. They evaluated their approach on the boundary layer flows to reduce computational costs and pointed out the necessity to develop tractable UQ approaches for computationally expensive cases.
Oliver and Moser~\cite{oliver2011bayesian} further extended the work of Cheung et al.~\cite{cheung2011bayesian} by introducing stochastic representations for the uncertainties in eddy viscosity turbulence models. The uncertainty representations based on the multiplicative error in mean velocity and the additive error in Reynolds shear stress are developed and used for plane channel flows.
Edeling et al.~\cite{edeling2014bayesian} proposed a Bayesian model-scenario averaging (BMSA) method to estimate the~$k$--$\epsilon$ turbulence model error for a class of boundary layer flows with different pressure gradients.
More recently, Edeling et al.~\cite{edeling2018bayesian} leveraged maximum a posteriori (MAP) estimate to reduce the computational cost and thus make their BMSA approach applicable for complex flows.

The aforementioned works use the Markov chain Monte Carlo (MCMC) technique which typically requires samples of at least $\mathcal{O}(10^5)$ to $\mathcal{O}(10^6)$. However, it would computationally intractable to deal with the complex flow cases of engineering interests where uncertainty propagation through the forward model is computationally expensive.
In order to reduce the computational cost, the conventional approach is to use surrogate models (e.g., the polynomial chaos methods~\cite{ma2009adaptive, edeling2016simplex, zhang2018efficient}) to replace the CFD code. 
Nevertheless, such approaches are challenging for high dimension problems due to the curse of dimensionality.
The ensemble technique has been proposed and discussed extensively for UQ problems in the data assimilation community. It can significantly reduce sample size to $\mathcal{O}(10^2)$ and provide reasonable estimates of posterior uncertainty with limited samples. Therefore, the ensemble methods can potentially play a role as an approximate Bayesian UQ approach for computationally expensive flow cases. The ensemble-based data assimilation methods will be further discussed below.

\subsection{Ensemble-based data assimilation}
Ensemble-based data assimilation has recently increased in popularity and has been applied to diverse contexts including fluid mechanics~\cite{zhang2019bayesian}, weather forecasting~\cite{liu2008ensemble} and geoscience~\cite{emerick2012history} due to its non-intrusiveness and robustness.
Among ensemble-based data assimilation methods, the most widely used is the ensemble Kalman filter (EnKF)~\cite{evensen2009data}.
It has been extensively used for uncertainty quantification in various applications, such as hydrology~\cite{gao2005quantifying,vrugt2007treatment}, meteorology~\cite{lorenc_potential_2003,houtekamer_review_2016}, oceangraphy~\cite{natvik_assimilation_2003,haugen_assimilation_2002}.
In the past few years, EnKF has also been increasingly leveraged for CFD applications to estimate empirical parameters or functional errors in the RANS closure models.
Kato and Obayashi~\cite{kato2013approach} explored the applicability of the EnKF method to estimate the uncertainty in the empirical parameters of the Spalart--Allmaras RANS model.
However, due to the strong nonlinearity of the RANS problem, it is necessary to iteratively assimilate data even for the stationary scenario, thus enhancing the performance of data fitting. To this end, Iglesias et al.~\cite{iglesias2013ensemble} proposed an iterative form of the standard EnKF as a derivative-free optimization method for inverse problems. In their framework, the analysis step of EnKF iterates with the artificial time for stationary systems based on the state augmentation.
They showed the accuracy of the iterative EnKF for inferring the sample mean with three different cases, but its accuracy in the context of uncertainty quantification has not been fully investigated.
Xiao et al.~\cite{xiao_quantifying_2016} applied this iterative EnKF to quantify and reduce the RANS model-form uncertainty within the Reynolds stress.
They demonstrated that the posterior mean with EnKF could have remarkably good agreement with benchmark data. 
The readers are referred to the recent review of Xiao and Cinnella~\cite{xiao2019quantification} for recent progress in model-form uncertainty quantification in RANS simulations.

For highly nonlinear systems, the ill-posedness of the problem is significantly increased. To search for the optimal point, EnKF takes a full gradient descent step where the forward model is linearized to simplify the problem~\cite{evensen2018analysis}. 
That possibly changes the original nonlinear problem and leads to wrong solutions.
Considering this issue, several iterative ensemble methods have been proposed and discussed for UQ of nonlinear systems in the data assimilation community.
For instance, Gu and Oliver~\cite{gu2007iterative} proposed the ensemble randomized maximum likelihood (EnRML) method to iterate the analysis step with the Gauss--Newton algorithm. They demonstrated the superiority of the EnRML method to EnKF for both static and dynamic problems with strong nonlinearity.
Chen and Oliver~\cite{chen2012ensemble} used the EnRML method as an iterative ensemble smoother for the history match problem.
Yang et al.~\cite{yang2015enhanced} proposed an enhanced ensemble variational method and applied their method to unsteady flows.
Their method is implemented similarly to EnRML with an iterative minimization procedure based on the Gauss--Newton algorithm, but the error covariance is updated sequentially based on ensemble analysis.
On the other hand, Emerick and Reynolds~\cite{emerick2012history} proposed an ensemble Kalman filter with multiple data assimilation (EnKF-MDA) and demonstrated it could provide better data match than EnKF with a comparable computational cost. This method performs Bayesian analysis with recursion of the likelihood through inflating the observation error.
It is worth noting that for unsteady cases, EnKF is usually used as a filtering technique to assimilate the data in time sequentially, while the EnRML method~\cite{chen2012ensemble} and EnKF-MDA~\cite{emerick2013ensemble} can be used as the smoother technique to account for all the available data simultaneously. 
Moreover, for the Gaussian linear case, it has been proven that the EnRML method and EnKF-MDA are equivalent to the EnKF~\cite{reynolds2006iterative,emerick2012history}. 
But for the nonlinear case, the equivalence does not hold. 
EnKF can be regarded as a single Gauss--Newton update with a full step. In contrast, the EnRML method and EnKF-MDA perform multiple small corrections, which helps to alleviate the inaccuracy caused by the linearization and better preserve the nonlinearity of the original problem.

The ensemble-based data assimilation methods mentioned above can be derived in a similar manner by solving the minimization problem under several mild assumptions (e.g., the Gaussian distribution, linearization, and ensemble gradient representation)~\cite{evensen2018analysis}.
However, these assumptions may result in a departure of the estimated posterior distribution from the truth. 
Recently, several authors investigated the cause of inaccurate uncertainty estimates given by the ensemble methods. 
For instance,
Oliver and Chen~\cite{oliver2011recent} reviewed the progress of MCMC, EnKF, and EnRML on the history matching problem. 
They concluded that the EnRML method could provide the probability distribution in better agreement with MCMC at a low computational cost, as compared to the EnKF method.
Ernst et al.~\cite{ernst2015analysis} examined the EnKF method for nonlinear stationary systems. They demonstrated that EnKF can provide the sample statistics as indication of uncertainties but is not suitable for rigorous Bayesian inference. 
Evensen~\cite{evensen2018analysis} derived and analyzed different ensemble methods from the view of model gradient representations and compared the analytic gradient and the ensemble representative gradient.
He concluded that none of these methods could provide the exact posterior probability density function (PDF) for highly nonlinear models, but they can serve as indication of the uncertainties at least for weakly nonlinear cases.
However, a sufficiently large number of samples is used to obtain accurate statistical estimation in his work, and the performance of these methods with small ensemble sizes is not fully evaluated. These iterative ensemble methods are useful for estimating uncertainties in QoIs in industrial CFD applications, and they warrant further investigation.

\subsection{Objective of present work}
In this work, we present the derivations of three different iterative ensemble methods, namely iterative EnKF~\cite{iglesias2013ensemble}(hereinafter referred to as EnKF for brevity), EnRML, and EnKF-MDA, from the optimization perspective, and compare their performances for quantifying uncertainties in steady-state CFD applications with small ensemble sizes. Moreover, the effect of small ensemble sizes on the performance of each method is evaluated in a scalar case by comparison with Bayesian distribution from MCMC.

The rest of the paper is structured as follows.
In Section~\ref{sec:derivation}, we give the brief derivation of the three most commonly used ensemble-based data assimilation methods (EnKF, EnRML, and EnKF-MDA). A scalar case is presented in Section~\ref{sec:scalar} to compare the performances of these methods with different ensemble sizes. In Section~\ref{sec:RANS}, a steady flow case is tested to identify the suitable approach to quantify the uncertainty in the RANS model. Section~\ref{sec:conclusion} concludes the paper.

\section{Ensemble-based data assimilation methods}
\label{sec:derivation}
Here, we summarize the brief derivation of the three different ensemble-based data assimilation methods (EnKF, EnRML, and EnKF-MDA) from the optimization perspective. For clarity and without loss of generality, we assume a multi-variate state-space model with multiple observations. This is in contrast to Evensen's work~\cite{evensen2018analysis} where a single-variate state-space model with a single observation is assumed.

\subsection{Minimization problem}
Consider that the observation model can be expressed as
\begin{equation}
	\mathsf{y} = \mathcal{H}[\mathsf{x}] + \epsilon,
\end{equation}
where~$\mathsf{x}$ is the state vector or input parameter~$\mathsf{x} \in \mathbb{R}^N$,~$\mathsf{y}$ is the observation~$\mathsf{y} \in \mathbb{R}^D$,~$\mathcal{H}$ is model function mapping the state to observation space~$\mathbb{R}^N \rightarrow \mathbb{R}^D$,
and~$\epsilon$ is added observation noise, which is assumed to be an independent and identically distributed (i.i.d.) Gaussian random vector with zero mean and covariance $\mathsf{R}$.
We give an initial guess on the PDF of state~$p(\mathsf{x})$ as the prior knowledge based on the Gaussian assumption.
Further, the Bayesian UQ approach can be used to find the posterior distribution conditioned by the observation.
The Bayes' theorem can be formulated as
\begin{equation}
    p(\mathsf{x}\mid\mathsf{y})\propto p(\mathsf{x}) \, p(\mathsf{y}\mid\mathcal{H}[\mathsf{x}]) \text{,}
    \label{eq:Bayes}
\end{equation}
which states that the posterior distribution~$p(\mathsf{x} \mid \mathsf{y})$ is proportional to the multiplication of the prior distribution~$p(\mathsf{x})$ and likelihood function~$p(\mathsf{y} \mid \mathcal{H}[\mathsf{x}])$ of data~$\mathsf{y}$ conditioned by the model~$\mathcal{H}[\mathsf{x}]$.

With the Gaussian assumption for prior~$p(\mathsf{x})$ and likelihood~$p(\mathsf{y}\mid \mathcal{H}[\mathsf{x}])$, we can rewrite the Bayes' formula in Eq.~\ref{eq:Bayes} as
\begin{equation}
    p(\mathsf{x} \mid \mathsf{y}) \propto p(\mathsf{x}) \, p(\mathsf{y}\mid \mathcal{H}[\mathsf{x}]) \propto e^{-J} ,
\end{equation}
where~$J$ is the cost function defined as
\begin{equation}
    J[\mathsf{x}^\text{a}] = \frac{1}{2}\left(\mathsf{x}^\text{a}-\mathsf{x}^\text{f}\right)^\top \mathsf{P}^{-1}\left(\mathsf{x}^\text{a}-\mathsf{x}^\text{f}\right)+\frac{1}{2}\left(\mathcal{H}[\mathsf{x}^\text{a}]-\mathsf{y}\right)^\top \mathsf{R}^{-1}\left(\mathcal{H}[\mathsf{x}^\text{a}]-\mathsf{y}\right).
\end{equation}
In the formula above,~$\mathsf{P}$ is the model error covariance,~$\mathsf{R}$ is the observation error covariance, and the superscripts~$\text{a}$ and~$\text{f}$ represent the ``analysis'' and ``forecast'', respectively.
It is challenging to obtain the true error covariance $\mathsf{P}$ in problems with high-dimensional state-spaces.
The ensemble methods apply the Monte Carlo technique to draw a small number of samples. Such samples can then be used to estimate the ensemble representation of the model error covariance~$\mathsf{P}$ and the observation error covariance~$\mathsf{R}$ as
\begin{equation}
\begin{aligned}
    \mathsf{P} &=\frac{1}{M-1}(\, \mathsf{X}-\bar{\mathsf{X}})\, (\mathsf{X}-\bar{\mathsf{X}} )^\top,
    \\\mathsf{R} &= \epsilon \epsilon^\top,
\end{aligned}
\end{equation}
where~$\mathsf{X}=\{\mathsf{x}_1, \ldots, \mathsf{x}_M \}$. Note that the estimated covariance matrix for the observation error and state are both symmetric.
Further, the maximum a posteriori (MAP) analysis can be applied to estimate the posterior distribution. That is, maximizing the posterior is equivalent to minimizing the cost function~$J$. 
Based on such an optimization perspective, we can derive the three different data assimilation methods, namely EnKF, EnRML, and EnKF-MDA, from the perspective of minimizing the cost function with different gradient descent techniques.

\subsection{EnKF}
\label{sec:EnKF}
For steady cases, the traditional EnKF only performs the Kalman update once.
It may be difficult to achieve a satisfactory data fit in some scenarios, for instance, where the prior mean is far from observation data, and the system model is strongly nonlinear~\cite{iglesias2013ensemble}.
To this end, the iterative technique is usually leveraged to adequately assimilate the data and thus prompt the data match.
We use an iterative form of EnKF proposed by Iglesias et al.~\cite{iglesias2013ensemble} to enhance the optimization performance. 
This method considers the EnKF as a regularized least square technique and performs multiple standard Kalman updates sequentially, even for stationary cases.
The cost function for each ensemble realization can be written as
\begin{equation}
    J[\mathsf{x}_{n,j}^\text{a}] = \frac{1}{2}\left(\mathsf{x}_{n,j}^\text{a}-\mathsf{x}_{n,j}^\text{f}\right)^\top \mathsf{P}_n^{-1}\left(\mathsf{x}_{n,j}^\text{a}-\mathsf{x}_{n,j}^\text{f}\right)+\frac{1}{2}\left(\mathcal{H}[\mathsf{x}_{n,j}^\text{a}]-\mathsf{y}\right)^\top \mathsf{R}^{-1}\left(\mathcal{H}[\mathsf{x}_{n,j}^\text{a}]-\mathsf{y}\right),
    \label{eq:cost_function}
\end{equation}
where~$n$ indicates the iteration number and~$j$ denotes the sample index.
Based on the cost function~\eqref{eq:cost_function}, the gradient with respect to the state is
\begin{equation}
    \frac{\partial J}{\partial \mathsf{x}_{n,j}^\text{a}}=\mathsf{P}_n^{-1}\left(\mathsf{x}_{n,j}^\text{a}-\mathsf{x}_{n,j}^\text{f}\right)+\left(\mathcal{H'}[\mathsf{x}_{n,j}^\text{a}]\right)^\top \mathsf{R}^{-1}\left(\mathcal{H}[\mathsf{x}_{n,j}^\text{a}]-\mathsf{y}_j\right) ,
    \label{eq:grad_costfunction}
\end{equation}
which should vanish to minimize the cost function $J$.
Therefore, the formulation of EnKF can be derived by setting the gradient of cost function~\eqref{eq:grad_costfunction} to be zero, which amounts to:
\begin{equation}
    \label{eq:zero_grad1_EnKF}
    \mathsf{P}_n^{-1}\left(\mathsf{x}_{n,j}^\text{a}-\mathsf{x}_{n,j}^\text{f}\right) = -\left(\mathcal{H'}[\mathsf{x}_{n,j}^\text{a}]\right)^\top \mathsf{R}^{-1}\left(\mathcal{H}[\mathsf{x}_{n,j}^\text{a}]-\mathsf{y}_j\right),
\end{equation}
where only the terms~$\mathcal{H'}[\mathsf{x}_j^\text{a}])$ and $\mathcal{H}[\mathsf{x}_j^\text{a}]$ are unknown. The assumption of linearization is introduced to have an estimation on the two unknown terms. The two terms are linearized as
\begin{subequations}
\begin{align}
     \label{eq:linear_assumption1}
     &\mathcal{H}[\mathsf{x}_j^\text{a}] \approx \mathcal{H}[\mathsf{x}_j^\text{f}] + \mathcal{H'}[\mathsf{x}_j^\text{f}]\left(\mathsf{x}_j^\text{a}-\mathsf{x}_j^\text{f}\right),\\
     \label{eq:linear_assumption2}
     &\mathcal{H'}[\mathsf{x}_j^\text{a}] \approx \mathcal{H'}[\mathsf{x}_j^\text{f}] + \mathcal{H''}[\mathsf{x}_j^\text{f}]\left(\mathsf{x}_j^\text{a} - \mathsf{x}_j^\text{f}\right),
\end{align}
\label{eq:linear_assumption}
\end{subequations}
where the second derivative in Eq.~\eqref{eq:linear_assumption2} is neglected for simplicity, assuming the model is moderately nonlinear.
With ensemble techniques, the model in observation space is randomized around the mean value~$\mathcal{H}[\bar{\mathsf{x}}^\text{f}]$.
After expanding~$\mathcal{H}[\mathsf{x}]$ around the ensemble mean~$\mathcal{H}[\bar{\mathsf{X}}]$,
we can represent~$\mathcal{H}[\mathsf{x}_j^\text{f}]$ with the model function gradient as~\cite{evensen2018analysis}
\begin{subequations}
\label{eq:ensemble_model_gradient}
\begin{align}
	&\mathcal{H}[\mathsf{x}_j^\text{f}] \approx \mathcal{H}[\bar{\mathsf{X}}^\text{f}] + \mathcal{H'}[\mathsf{x}_j^\text{f}] \left(\mathsf{x}_j^\text{f}-\bar{\mathsf{X}}^\text{f}\right) \text{.}
\end{align}
\end{subequations}
We introduce the tangent linear model~$\mathcal{H}[\mathsf{x}] = \mathsf{Hx}$, and thus the gradient representation~$\mathcal{H'}[\mathsf{x}^\text{f}]$ can be expressed as the tangent linear operator~$\mathsf{H}$ by assuming the linear relationship between the measurement and the state.
Accordingly, the update step of EnKF can be derived and formulated as
\begin{equation}
    \label{eq:update_EnKF}
    \mathsf{x}_{n,j}^\text{a} = \mathsf{x}_{n,j}^\text{f}+ \mathsf{P}_n \mathsf{H}^\top \left( \mathsf{R} + \mathsf{H P}_n \mathsf{H}^\top\right)^{-1}\left(\mathsf{y}_j-\mathsf{H} \mathsf{x}_{n,j}^\text{f}\right).
\end{equation}
Due to practical consideration, one does not usually compute the model operator~$\mathsf{H}$ explicitly. Rather, the two terms~$\mathsf{PH}^\top$ and~$\mathsf{HPH}^\top$ can be reformulated as
\begin{subequations}
\begin{align}
    \mathsf{P H}^\top &= \frac{1}{M-1}\left(\mathsf{X}-\bar{\mathsf{X}}\right)\left(\mathcal{H}[\mathsf{X}]-\mathcal{H}[\bar{\mathsf{X}}]\right)^\top,\\ 
    \mathsf{HPH}^\top &= \frac{1}{M-1}\left(\mathcal{H}[\mathsf{X}]-\mathcal{H}[\bar{\mathsf{X}}]\right)\left(\mathcal{H}[\mathsf{X}]-\mathcal{H}[\bar{\mathsf{X}}]\right)^\top.
\end{align}
\end{subequations}
Besides, the ensemble observation is adopted based on~\cite{burgers1998analysis}. 
That is,  we use randomly perturbed observation data for each realization.
Further details of the derivation are presented in~\ref{appendixA}.
We emphasize that the iterative ensemble Kalman method is a specific method for solving inverse problems that is distinct from the conventional EnKF.
It regards the ensemble Kalman method as the regularized least square technique.
For stationary problems, the update step is iterated with pseudo-time to reduce data misfit.
The iterative ensemble Kalman method for uncertainty quantification will be further discussed in subsection~\ref{sec:remark}.

\subsection{EnRML}
The ensemble randomized maximum likelihood method~\cite{gu2007iterative} updates the initial guess of state vector iteratively with Gauss--Newton algorithm. The cost function can be written as
\begin{equation}
    J[\mathsf{x}_{l,j}]=\frac{1}{2}\left(\mathsf{x}_{l,j}-\mathsf{x}_{0,j}\right)^\top \mathsf{P}_0^{-1}\left(\mathsf{x}_{l,j}-\mathsf{x}_{0,j}\right)+\frac{1}{2}\left(\mathcal{H}{[\mathsf{x}_{l,j}]}-\mathsf{y}_j\right)^\top \mathsf{R}^{-1} \left(\mathcal{H}{[\mathsf{x}_{l,j}]}-\mathsf{y}_j\right),
    \label{eq:costfunction_EnRML}
\end{equation}
where~$\mathsf{x}_0$ is the initial guess,~$\mathsf{P}_0$ is the initially estimated model error covariance before the data assimilation process~,~and iteration index $l$ indicates the sub-iteration of the EnRML method.
The gradient and Hessian of the cost function~\eqref{eq:costfunction_EnRML} can be derived similarly as in EnKF
\begin{subequations}
\begin{align}
    \label{eq:grad_EnRML}
    & \frac{\partial J}{\partial \mathsf{x}_{l,j}}=\mathsf{P}_0^{-1}\left(\mathsf{x}_{l,j}-\mathsf{x}_{0,j}\right)+\mathcal{H}'[\mathsf{x}_{l,j}]^\top \mathsf{R}^{-1}\left(\mathcal{H} [\mathsf{x}_{l,j}]-\mathsf{y}_j\right),\\
    \label{eq:hes_EnRML}
    & \frac{\partial^2 J}{\partial^2 \mathsf{x}_{l,j}} = \mathsf{P}_0^{-1}+ \mathcal{H}' [\mathsf{x}_{l,j}]^\top \mathsf{R}^{-1} \mathcal{H}' [\mathsf{x}_{l,j}].
\end{align}
\end{subequations}
Instead of reaching a zero-gradient minimum directly as in EnKF, the prior~$\mathsf{x}_0$ is iteratively updated based on Gauss--Newton method as
\begin{equation}
    \label{eq:Gauss_Newton}
    \mathsf{x}_{l,j}^\text{a} = \mathsf{x}_{l,j}^\text{f} - \gamma\left(\frac{\partial^2 J}{\partial \mathsf{x}_{l,j}^2}\right)^{-1}{\frac{\partial J}{\partial \mathsf{x}_{l,j}}},
\end{equation}
where~$\gamma$ is the step length parameter. The Gauss--Newton approach can reduce the step length and ease the influence of the linearization assumption during the analysis step.
With the gradient~\eqref{eq:grad_EnRML} and the Hessian~\eqref{eq:hes_EnRML} of the cost function we can obtain the analysis scheme for the EnRML method as follows:
\begin{equation}
\begin{aligned}
    \label{eq:update_EnRML}
    \mathsf{x}_{l,j}^\text{a} = \gamma \mathsf{x}_{0,j}^\text{f} +\left(1- \gamma\right) \mathsf{x}_{l,j}^\text{f} - &\gamma \mathsf{P}_0 \mathcal{H}' [\mathsf{x}_{l,j}^\text{f}]^\top \left(\mathsf{R}+\mathcal{H}'[\mathsf{x}_{l,j}^\text{f}]^\top \mathsf{P}_0 \mathcal{H}' [\mathsf{x}_{l,j}^\text{f}]\right)^{-1}\\
    &\left(\mathcal{H}[\mathsf{x}_{l,j}^\text{f}]-\mathsf{y}_j-\mathcal{H}'[\mathsf{x}_{l,j}^\text{f}]\left(\mathsf{x}_{l,j}^\text{f}-\mathsf{x}_{0,j}^\text{f}\right)\right).
\end{aligned}
\end{equation}
In the EnRML method, the model error covariance~$\mathsf{P}$ remains as the initial one $\mathsf{P}_0$ and does not change with the iteration number. Moreover, the sensitivity matrix~$\mathcal{H}'[\mathsf{X}]$ has to be evaluated at each iteration through
\begin{equation}
    \label{eq:ensemble_assumption_EnRML}
    \mathcal{H}'[\mathsf{X}_l] \approx \left(\mathcal{H}[\mathsf{X}_l] - \mathcal{H}[\bar{\mathsf{X}}_l]\right)\left(\mathsf{X}_l-\bar{\mathsf{X}}_l\right)^{-1} \text{.}
\end{equation}
The singular value decomposition~(SVD) is used to estimate the inverse of the non-full rank matrix. The details of the derivation can be found in~\ref{appendixB}.

\subsection{EnKF-MDA}
From the derivation above, each update of EnKF can be regarded as the Gauss--Newton update but uses a full step in the search direction. However, a single global update may not result in a satisfactory data fit.
Hence, assimilating the data multiple times is highly desired to improve the data fit~\cite{reynolds2006iterative}.
Moreover, in some cases where the prior mean/first guess is far from the truth and the model is highly nonlinear, performing the full Gauss--Newton step may result in the overcorrection and lead to inaccurate solutions. This deficiency can be alleviated to damp the changes in the early iterations~\cite{wu1998conditioning,gao2004improved}.
To this end, Emerick and Reynolds~\cite{emerick2012history} proposed EnKF-MDA to assimilate the same data multiple times with an inflated observation error covariance. 
They have proven that for linear Gaussian cases, the EnKF-MDA is equivalent to the EnKF. For nonlinear cases, the traditional EnKF uses a full Gauss--Newton step with an average sensitivity estimated from the prior ensemble and probably leads to a large Gauss--Newton correction~\cite{reynolds2006iterative}. EnKF-MDA can be regarded as performing multiple small corrections to damp the changes of the model and thus alleviate the effects of nonlinearity~\cite{emerick2013ensemble}.
From the Bayesian perspective, the likelihood function of EnKF-MDA is in a recursive form as
\begin{equation}
    p(\mathsf{x}\mid \mathsf{y}) \propto p(\mathsf{x}) \prod_{l=1}^{N_\text{mda}} p(\mathsf{y}\mid \mathcal{H}[\mathsf{x}_{l-1}])^{\frac{1}{\alpha_l}},
\end{equation}
where~$\sum_{l=1}^{N_\text{mda}}\frac{1}{\alpha_l}=1$,~$N_\text{mda}$ is the total data assimilation iteration number, and~$\alpha$ can be chosen simply as $N_\text{mda}$. The cost function~$J$ can be expressed as:
\begin{equation}
    \label{costfunction_EnKF_MDA}
    J[\mathsf{x}_{l,j}^\text{a}]=\frac{1}{2}\left(\mathsf{x}_{l,j}^\text{a}-\mathsf{x}_{l,j}^\text{f}\right)^\top \mathsf{P}_l^{-1}\left(\mathsf{x}_{l,j}^\text{a}-\mathsf{x}_{l,j}^\text{f}\right)+ \frac{1}{2}\left(d+\sqrt{\alpha_l}\epsilon_{l,j} - \mathcal{H}[\mathsf{x}_{l,j}^\text{a}]\right)^\top \left(\alpha_l \mathsf{R}\right)^{-1} \left(d+\sqrt{\alpha_l}\epsilon_{l,j}-\mathcal{H}[\mathsf{x}_{l,j}^\text{a}]\right) ,
\end{equation}
where $d$ is the measurement without perturbations. The gradient of the cost function is then
\begin{equation}
    \label{eq:grad_EnKF_MDA}
    \frac{\partial J[\mathsf{x}_{l,j}^\text{a}]}{\partial \mathsf{x}_{l,j}^\text{a}}=\mathsf{P}_l^{-1}\left(\mathsf{x}_{l,j}^\text{a}-\mathsf{x}_{l,j}^\text{f}\right)+\mathcal{H'}[\mathsf{x}_{l,j}^\text{a}]^\top \left(\alpha_l \mathsf{R}\right)^{-1}\left(d+\sqrt{\alpha_l}\epsilon_{l,j}- \mathcal{H}[\mathsf{x}_{l,j}^\text{a}]\right).
\end{equation}
Similar to the derivation of EnKF method, we set the gradient of cost function to zero.
Further, with the linearization assumption~\eqref{eq:linear_assumption} and ensemble gradient representation~\eqref{eq:ensemble_model_gradient}, we have the update scheme as
\begin{equation}
    \label{eq:update_EnKF_MDA_int}
    \mathsf{x}_{l,j}^\text{a} = \mathsf{x}_{l,j}^\text{f} + \mathsf{P}_l \mathcal{H'}[\mathsf{x}_{l,j}^\text{f}]^\top \left(\mathcal{H}[\mathsf{x}_{l,j}^\text{f}] \mathsf{P}_l \mathcal{H}[\mathsf{x}_{l,j}^\text{f}]^\top + \alpha_l \mathsf{R}\right)^{-1}\left(d+\sqrt{\alpha_l}\epsilon_{l,j}-\mathcal{H}[\mathsf{x}_{l,j}^\text{f}]\right).
\end{equation}
By introducing the tangent linear operator~$\mathsf{H}$, we can obtain the analysis step of EnKF-MDA as
\begin{equation}
    \label{eq:update_EnKF_MDA}
    \mathsf{x}_{l,j}^\text{a} = \mathsf{x}_{l,j}^\text{f} + \mathsf{P}_l \mathsf{H}^\top \left(\mathsf{H P}_l \mathsf{H}^\top+\alpha_l \mathsf{R}\right)^{-1}\left(d+\sqrt{\alpha_l}\epsilon_{l,j}-\mathsf{H x}_{l,j}^\text{f}\right) \text{.}
\end{equation}

Given the prior distribution of system state and ensemble observations with error covariance matrix~$\mathsf{R}$, the implementation for the three data assimilation methods are summarized in Table~\ref{tab:scheme_comparison}.
\begin{table}[!htbp]
\centering
\begin{tabular}{l | l | l}

\hline
\qquad EnKF & \qquad EnKF-MDA & \qquad EnRML \\
\hline
\multicolumn{2}{p{8cm}|}{\textbf{a. sampling step:}}
& \multicolumn{1}{p{4cm}}{\textbf{a. sampling step:}}
\\
\multicolumn{2}{l|}{ \quad generate initial ensemble state vectors ${\{\mathsf{x}_{0,j}}\}_{j=1}^M$}
& \multicolumn{1}{p{6cm}}{\quad \emph{1.} generate initial ensemble state vectors ${\{\mathsf{x}_{0,j}}\}_{j=1}^M$;}\\
\multicolumn{2}{l|}{}
&  \multicolumn{1}{p{6cm}}{\quad \emph{2.} estimate the mean $\bar{\mathsf{X}}_0^\text{f}$ and model error covariance $\mathsf{P}_0$ of the ensemble.}\\
\hline
\multicolumn{2}{p{8cm}|}{\textbf{b. prediction step:}} 
& \multicolumn{1}{p{6cm}}{\textbf{b. prediction step:}} \\
\multicolumn{2}{p{8cm}|}{ \quad i) Propagate from current state $l-1$ to next iteration level $l$ based on forward model $(l>0)$.}
& \multicolumn{1}{p{6cm}}{ \quad i) Propagate from current state $l-1$ to next iteration level $l$ based on forward model $(l>0)$.}
\\
\multicolumn{2}{c|}{$\mathsf{x}_{l,j}^\text{f} = \mathcal{F}[\mathsf{x}_{l-1,j}^\text{a}]$}
& \multicolumn{1}{c}{$\mathsf{x}_{l,j}^\text{f} = \mathcal{F}[\mathsf{x}_{l-1,j}^\text{a}]$}
\\
\multicolumn{2}{p{8cm}|}{ \quad ii) Estimate the ensemble mean $\bar{\mathsf{X}}_l^\text{f}$ and model error covariance $\mathsf{P}_l$ of the current iteration.}
&  \multicolumn{1}{p{6cm}}{ \quad ii) Estimate the ensemble model gradient by \eqref{eq:ensemble_assumption_EnRML}.}\\
\hline
\textbf{c. analysis step} & \textbf{c. analysis step} & \textbf{c. analysis step}\\
\multicolumn{1}{p{4cm}|}{update the state vector by \eqref{eq:update_EnKF} and return to step b until the convergence criteria are reached.}
& \multicolumn{1}{p{4cm}|}{update the state vector by \eqref{eq:update_EnKF_MDA} and return to step b until the convergence criteria are reached.}
& \multicolumn{1}{p{6cm}}{update the state vector by \eqref{eq:update_EnRML} and return to step b until the convergence criteria are reached.}\\
\hline
\end{tabular}
\caption{Schematic comparison of EnKF, EnRML and EnKF-MDA}
\label{tab:scheme_comparison}
\end{table}

\subsection{Remarks}
\label{sec:remark}
From the derivations above, we apply the iterative form, linearization assumption, and ensemble gradient representation to obtain the derivative-free analysis scheme. Here, we provide some discussion on the effects of each issue.
\begin{enumerate}
\item Iterative form is necessary to obtain satisfactory inference results for the inverse problem of nonlinear systems. However, 
the iterative EnKF performs several Gauss--Newton iterations with the full step where data is equally used for stationary systems. While the other two methods conduct partial iterations, and the several sub-iterations are only equivalent to the first iteration of the iterative EnKF illustrated in Section~\ref{sec:EnKF}.
This iterative EnKF may cause the samples to collapse in early iterations and leads to underestimation of uncertainty, since the data is repeatedly used. 
Moreover, as the model error covariance for the next iteration becomes very small, the first term in the cost function~\eqref{eq:cost_function} prescribing the prior distribution will be dominant. 
That means the data assimilation analysis does not take effect, and the update only depends on the prior afterward. 
In contrast, the EnRML method and EnKF-MDA iterate the update step through the Gauss--Newton algorithm and likelihood recursion, respectively, which can avoid the data overuse and sample collapse.
\item The linearization assumption is introduced in our derivation for simplification. However, for strongly nonlinear systems, the linear assumption may significantly affect the optimal solution and lead to inaccurate inference results. EnKF takes a full update step to the optimal point, while the EnRML method and EnKF-MDA split one EnKF step by several small steps through Gauss--Newton method and likelihood recursion, respectively. From this perspective, the EnRML method and EnKF-MDA can alleviate the influence of linearization and partly preserve the nonlinearity.
Therefore, the EnRML method and EnKF-MDA are more suitable for the uncertainty quantification of stationary nonlinear systems than the iterative EnKF.
\item Another assumption, ensemble gradient representation, is leveraged in the ensemble-based DA methods as presented in our derivations. That is, the model gradient is approximated by ensemble realizations and is not derived analytically. This may cause the propagated posterior distribution to depart from the exact Bayesian distribution~\cite{evensen2018analysis}. While the impact of linearization can be alleviated through the Gauss-Newton algorithm or reduced likelihood recursion, the effects of ensemble gradient representation are inevitable for ensemble methods unless the adjoint method is used to compute the analytic gradient.
\end{enumerate}
Moreover, the parameters~$\gamma$ and~$N_\text{mda}$ which control the length of the update step are introduced in the EnRML method and EnKF-MDA, respectively. They can be constant or adaptive based on the convergence judgment. Specifically, if the discrepancy in observation space is larger than that in the last iteration, we can reduce the step length by decreasing the step length parameter $\gamma$ or increasing the inflation parameter $N_\text{mda}$. Conversely, if the discrepancy is reduced, we can increase the $\gamma$ in EnRML or reduce the $N_\text{mda}$ in EnKF-MDA to speed up the convergence~\cite{emerick2013ensemble}.

\section{Scalar case}
\label{sec:scalar}
We first test the three ensemble-based Bayesian UQ approaches derived in Section~2 on a simple case used by Evensen~\cite{evensen2018analysis}. In his work, the effects of the model gradient representation are investigated with a sufficiently large sample size.
Here, we focus on the effects of limited ensemble sizes and evaluate the performance of the iterative ensemble methods with small sample sizes.
In this case, the computing time for the forward model is negligible.
Hence, we can obtain Bayesian posterior from MCMC and ensemble methods with a large sample size for comparison.

\subsection{Problem statement}
The forward model is defined as:
\begin{equation}
    \hat{\mathsf{y}} = 1 + \sin (\pi \mathsf{x}) + q,
\end{equation}
where~$\mathsf{x}$ is the state variable,~$\hat{\mathsf{y}}$ is the model output in observation space, and~$q$ is the added model error with~$q \sim \mathcal{N}(0,0.03^2)$. The goal is to quantify and reduce the uncertainty of~$\mathsf{x}$ and~$\hat{\mathsf{y}}$ with Bayesian approaches.
The Bayesian UQ approach need the statistical information on the prior state and the observation data.
We assume that the state variable~$\mathsf{x}$ and data~$\mathsf{y}$ both obey to the Gaussian distribution as~$\mathsf{x} \sim \mathcal{N}(0,0.1^2)$ and~$\mathsf{y} \sim \mathcal{N}(1,0.1^2)$.
We set the step length parameter~$\gamma$ in the EnRML method as~$0.5$ and the inflation parameter~$N_{\text{mda}}$ in EnKF-MDA as~$30$ to obtain convergence results.
The performance of the ensemble methods is assessed with two different ensemble sizes of~$10^6$ and~$10^2$, and the effects of small ensemble sizes on the propagated uncertainties are investigated.
We conduct the Markov chain Monte Carlo (MCMC) with $10^8$ samples by using the DREAM algorithm~\cite{vrugt2016markov} and consider the results as the gold standard.
The probability density in this case is estimated from the samples through kernel density estimation (KDE) using the Gaussian kernel.

From the derivation in Section~2, it has been noted that two assumptions (linearization and ensemble gradient representation) are introduced to obtain the derivative-free analysis step. The model gradient can be represented by the analytic gradient or estimated by the ensemble samples. Although the analytic model gradient can give more accurate results compared to the ensemble gradient representation~\cite{evensen2018analysis}, it is not practical for complex models and beyond the scope of this work.
Here, we focus on the ensemble gradient and also investigate the effects of ensemble sizes on the ensemble gradient. The Python code for this test case is provided in a publicly available GitHub repository~\cite{scalar-git}.

\subsection{Results}
\begin{figure}[!htbp]
    \centering
    \begin{subfigure}[b]{0.45\linewidth}
    \includegraphics[width=\textwidth]{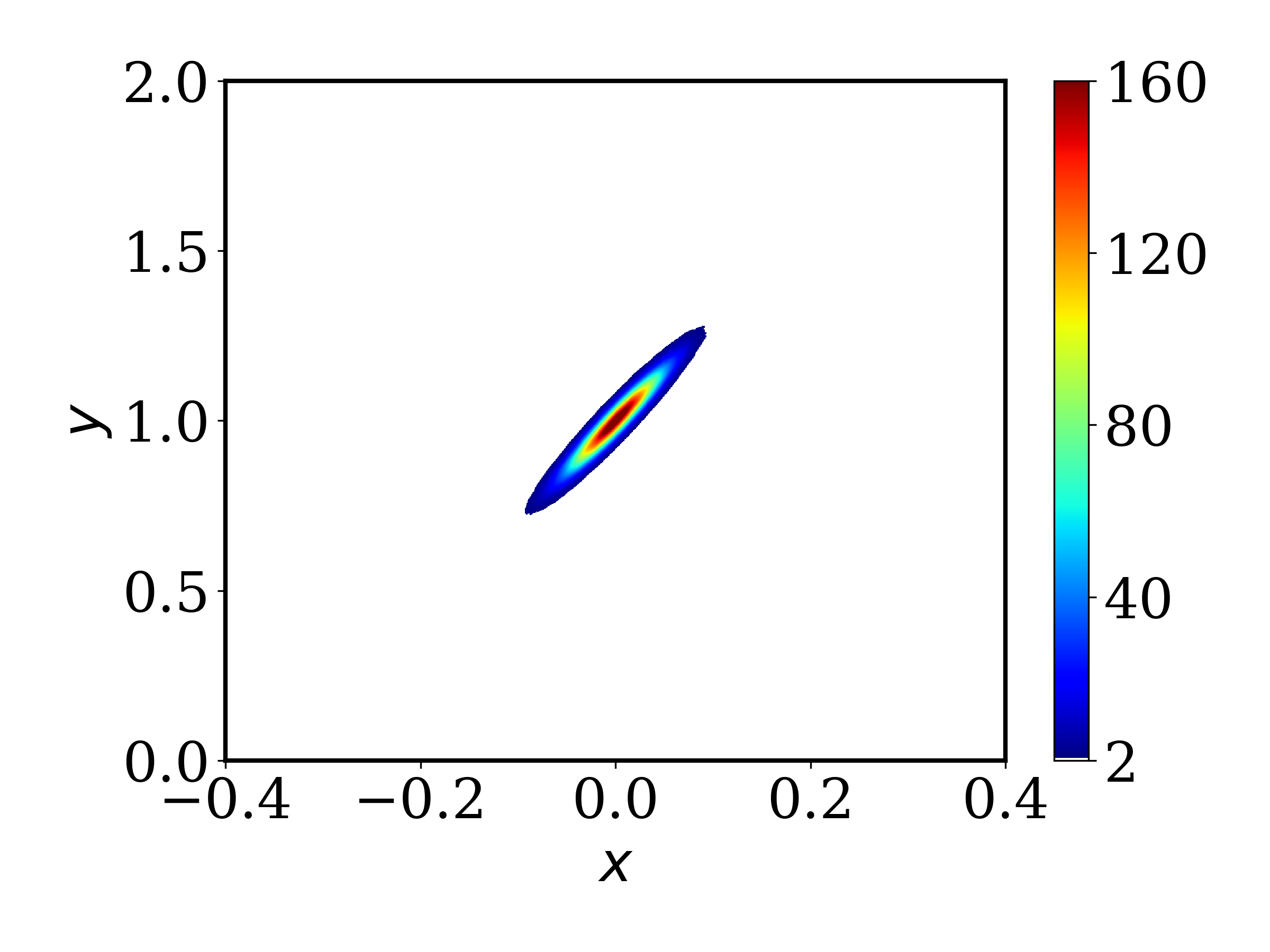}
    \caption{MCMC}
    \end{subfigure}
    \begin{subfigure}[b]{0.45\linewidth}
    \includegraphics[width=\textwidth]{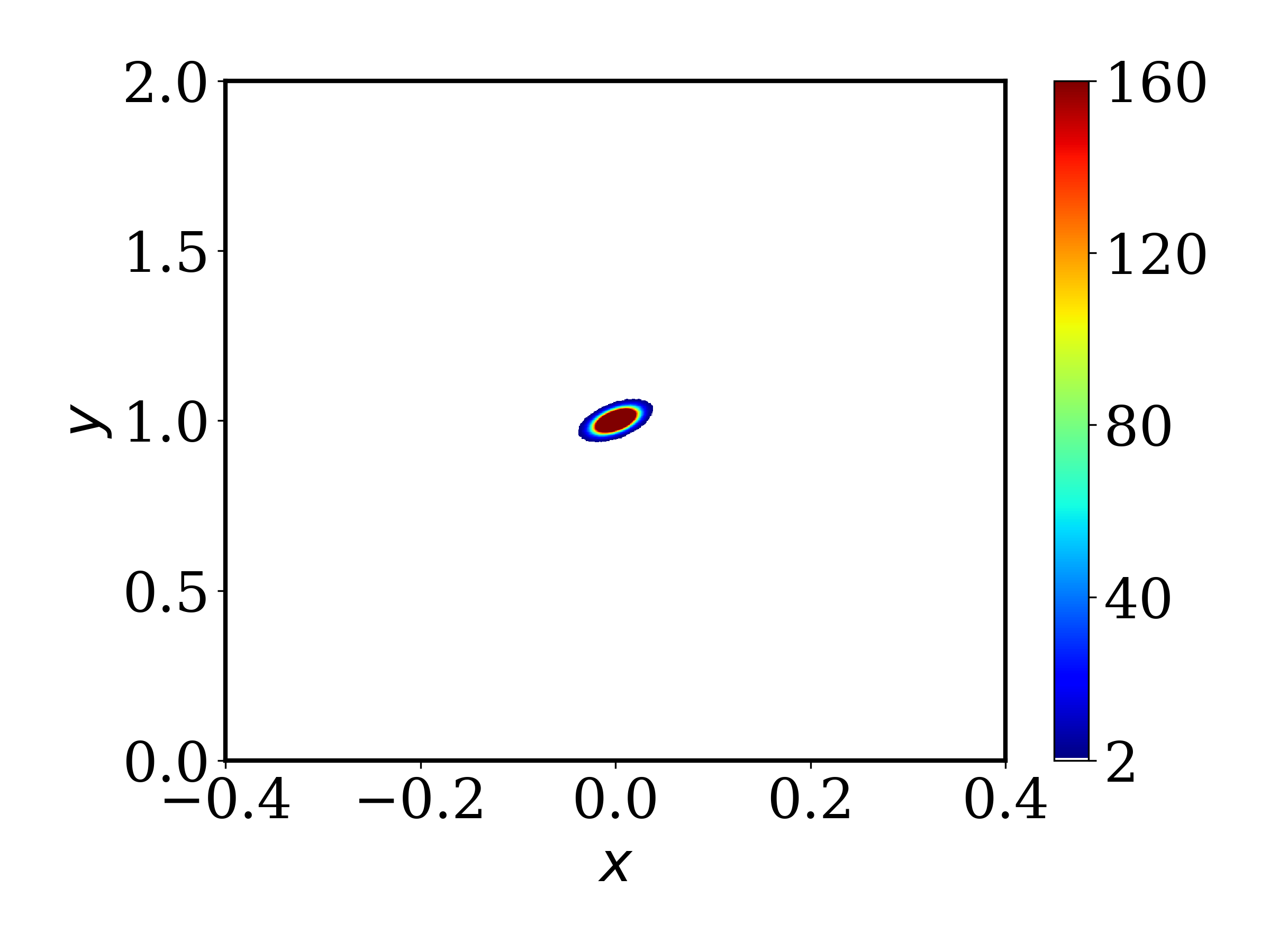}
    \caption{EnKF}
    \end{subfigure}
    \begin{subfigure}[b]{0.45\linewidth}
    \includegraphics[width=\textwidth]{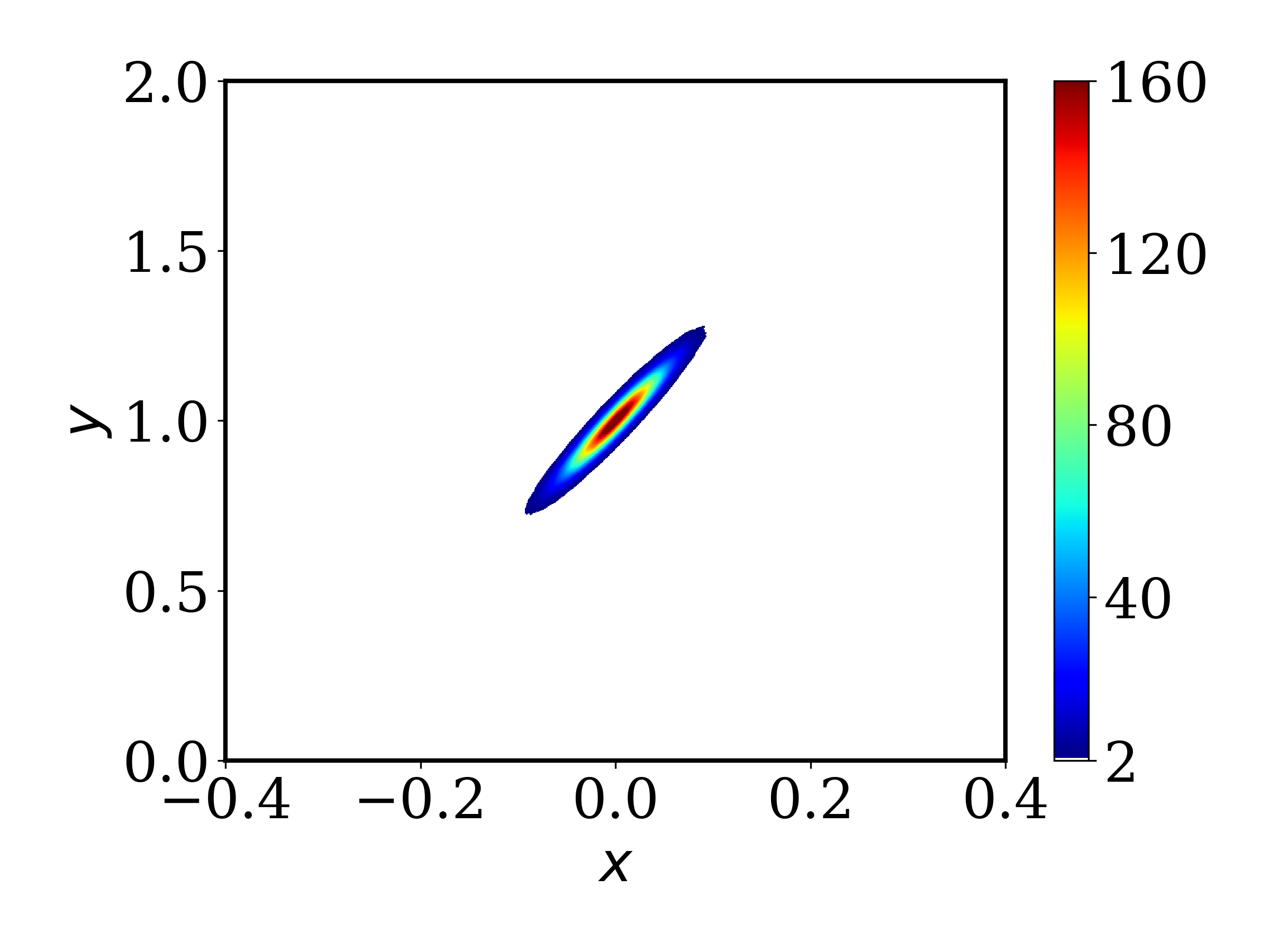}
    \caption{EnRML}
    \end{subfigure}
    \begin{subfigure}[b]{0.45\linewidth}
    \includegraphics[width=\textwidth]{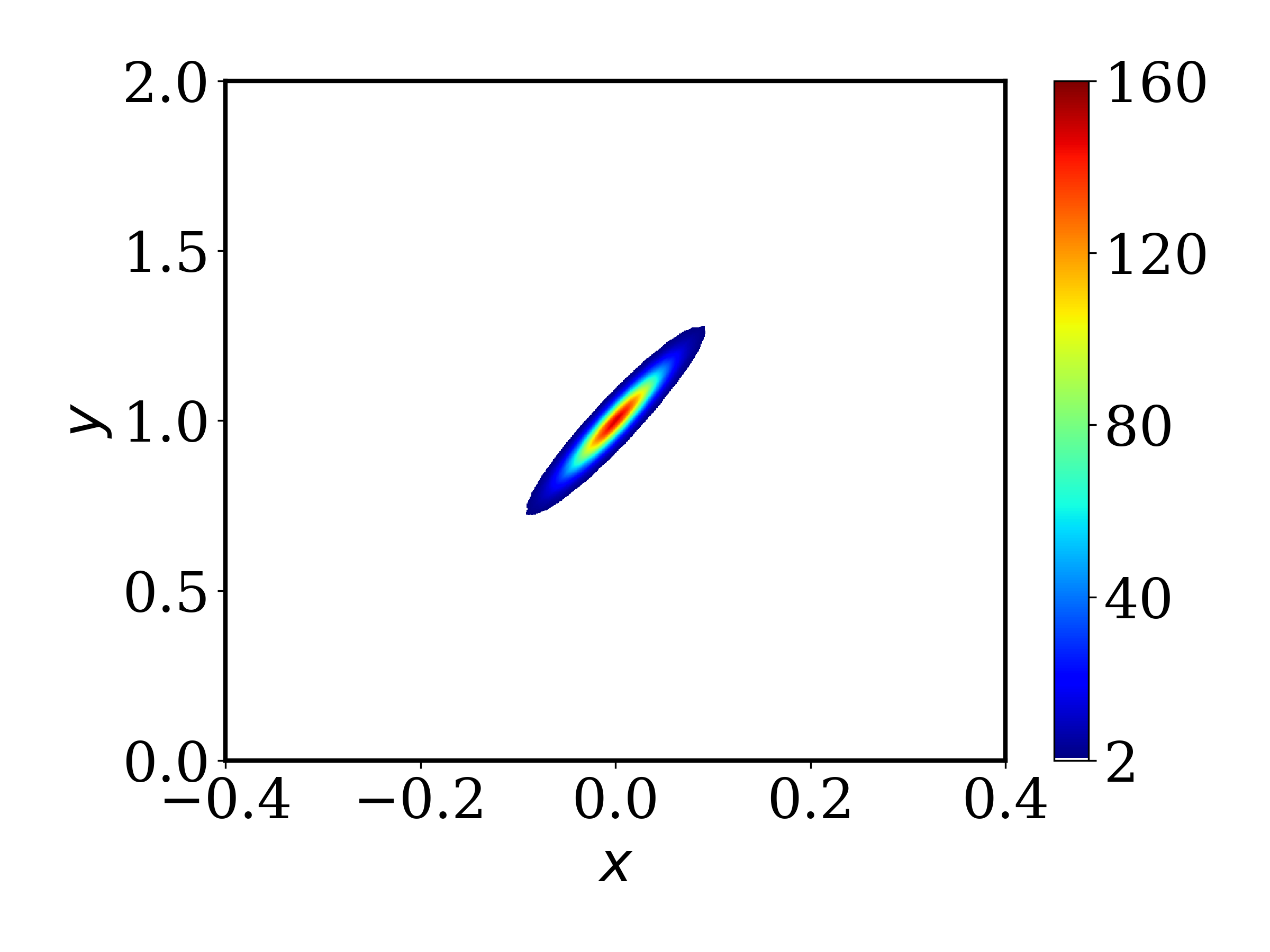}
    \caption{EnKF--MDA}
    \end{subfigure}
    \caption{Joint PDFs with $\bm{10^6}$ \textbf{samples} with the comparison among Bayes, EnKF, EnRML, and EnKF-MDA for the scalar case.}
    \label{fig:scalar_large_size}
\end{figure}

\begin{figure}[!htbp]
    \centering
    \includegraphics[width=0.8\textwidth]{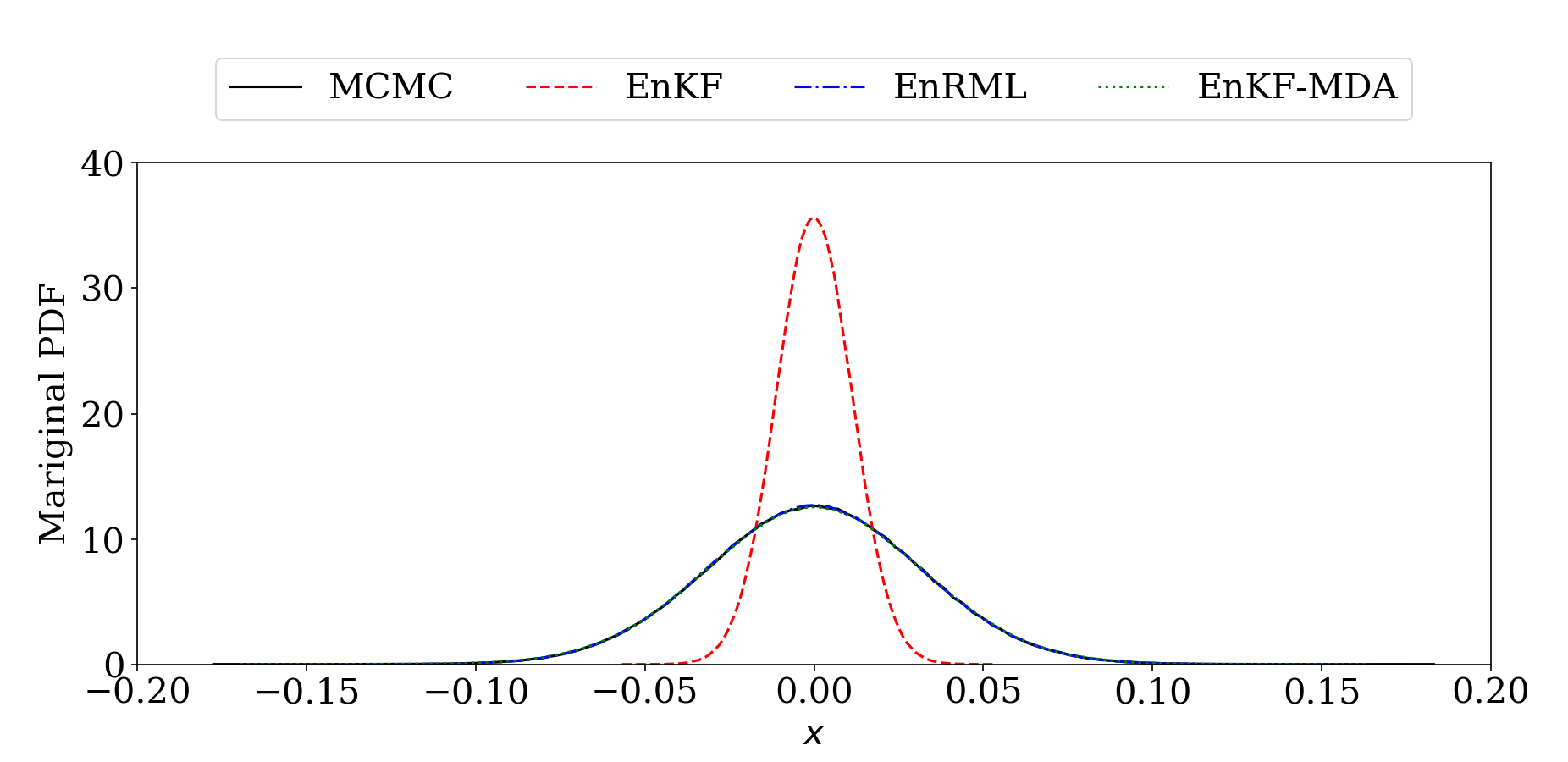}
    \caption{Marginal PDFs for $\mathsf{x}$ with $\bm{10^6}$ \textbf{samples} with the comparison among MCMC, EnKF, EnRML, and EnKF-MDA for the scalar case.}
    \label{fig:marginal_pdf_largesize}
\end{figure}

We first evaluate the performance of each ensemble method with a large ensemble size~$M = 10^6$.
The joint and marginal PDFs with comparison among different ensemble methods are shown in Fig.~\ref{fig:scalar_large_size} and Fig.~\ref{fig:marginal_pdf_largesize}, respectively.
From the results, it can be seen that all the three ensemble methods can capture the posterior mean. However, it is apparent that the iterative EnKF method leads to overconfidence in the mean value and significantly underestimates the posterior variance compared to the exact Bayesian distribution from MCMC. On the contrary, both the EnRML method and EnKF-MDA can provide an estimation on the posterior distribution in good agreement with the benchmark data. 
This is not surprising since the iterative EnKF repeats using the same data, while the EnRML method and EnKF-MDA can avoid data overuse by introducing the Gauss--Newton method or the observation error inflation, as we remarked in Section~2.
To summarize, with large ensemble size, the EnRML method and EnKF-MDA can perform comparably to the MCMC, while EnKF significantly underestimates the posterior uncertainty due to data repeatedly used.

\begin{figure}[!htbp]
    \centering
    \begin{subfigure}[b]{0.45\linewidth}
        \includegraphics[width=\textwidth]{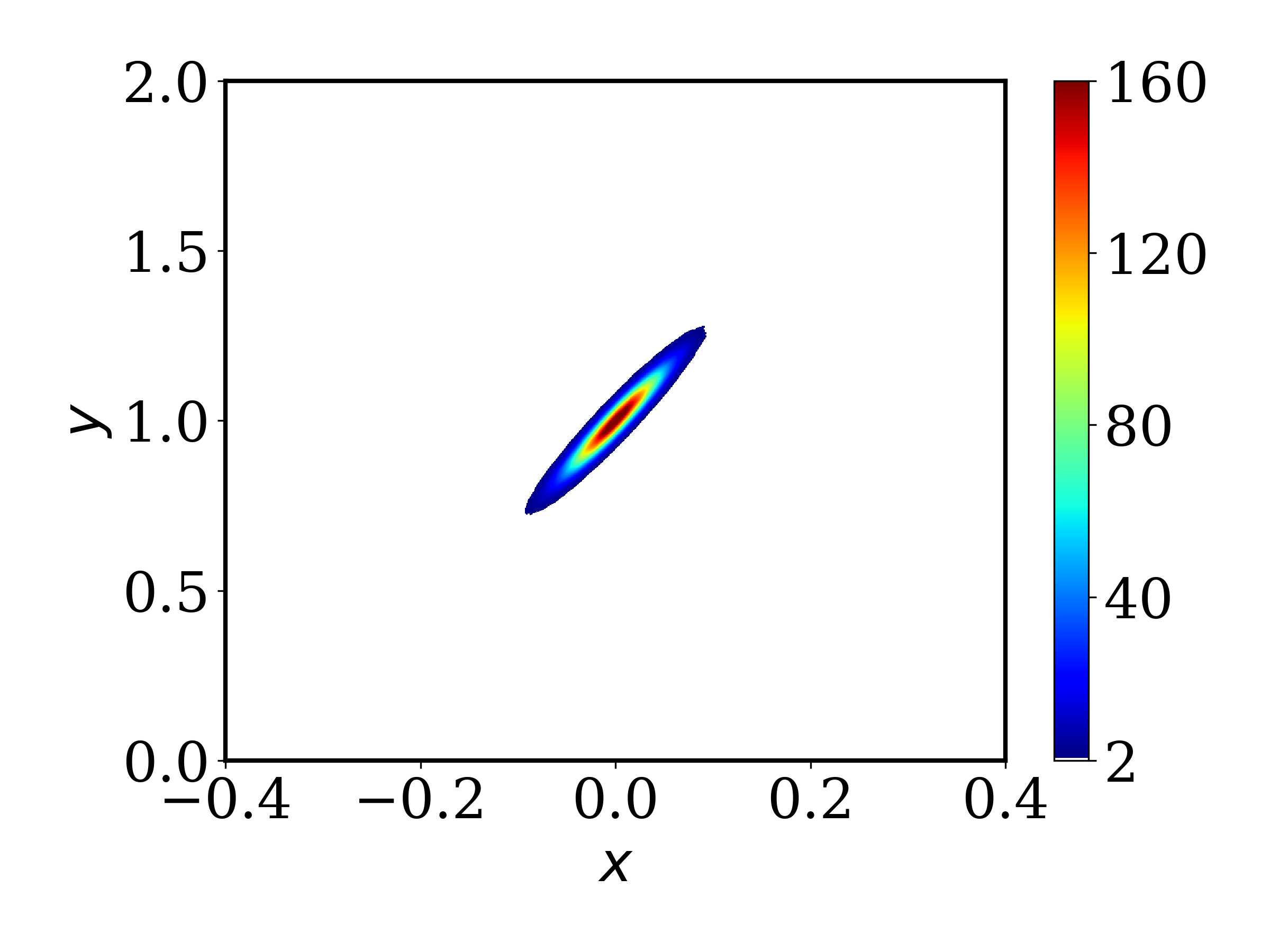}
        \caption{MCMC}
    \end{subfigure}
    \begin{subfigure}[b]{0.45\linewidth}
        \includegraphics[width=\textwidth]{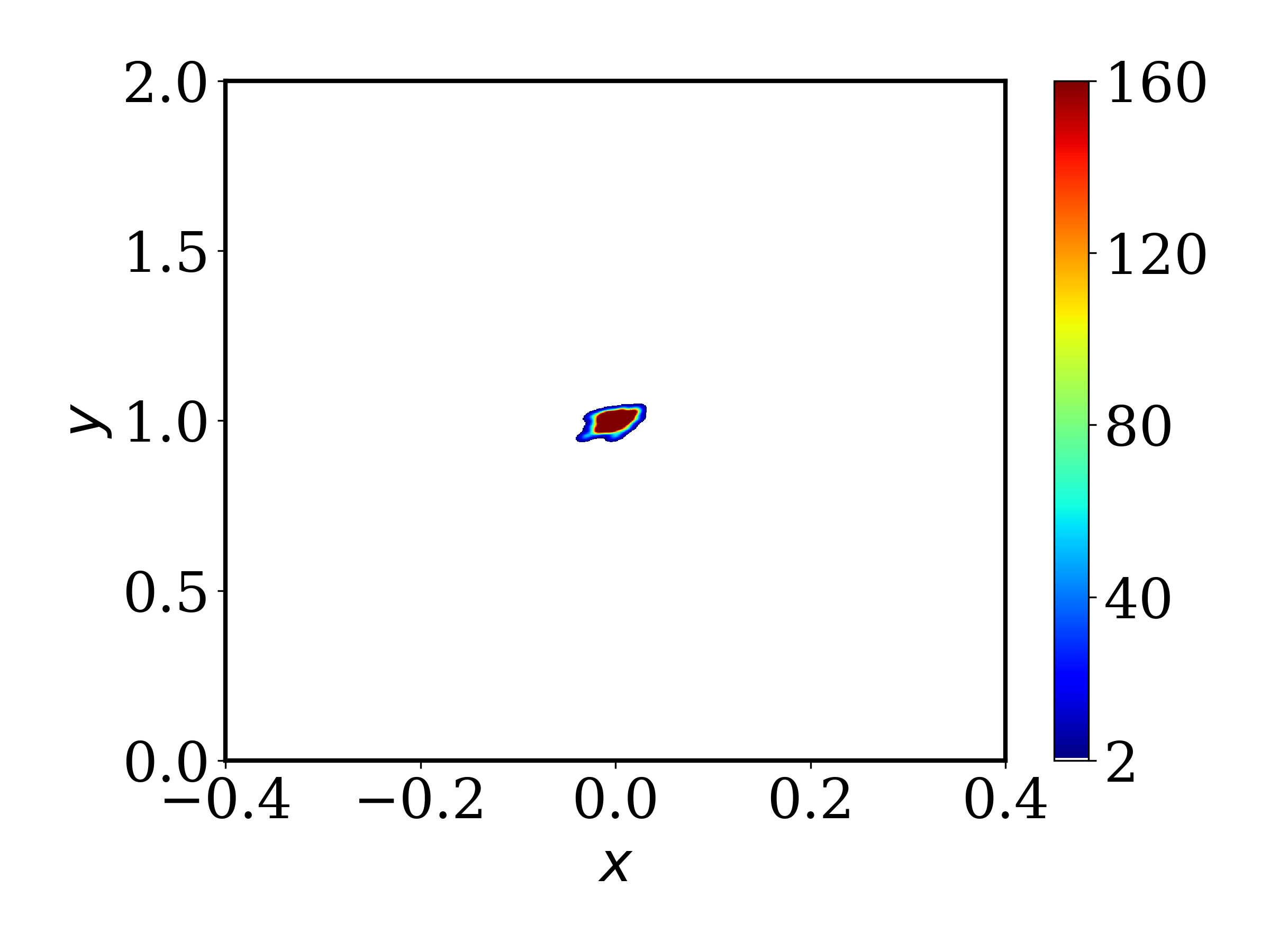}
        \caption{EnKF}
    \end{subfigure}
    \begin{subfigure}[b]{0.45\linewidth}
        \includegraphics[width=\textwidth]{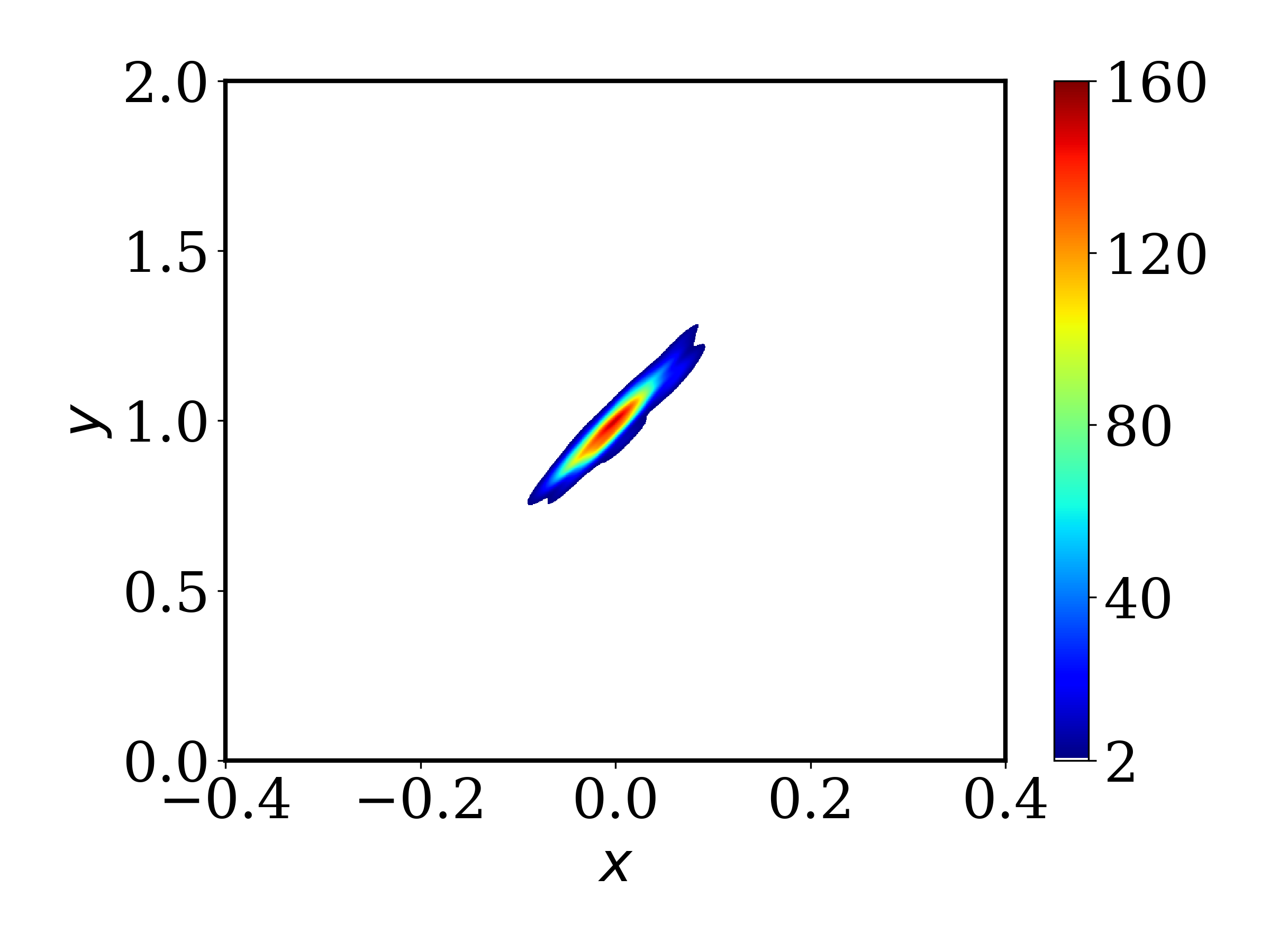}
        \caption{EnRML}
    \end{subfigure}
    \begin{subfigure}[b]{0.45\linewidth}
        \includegraphics[width=\textwidth]{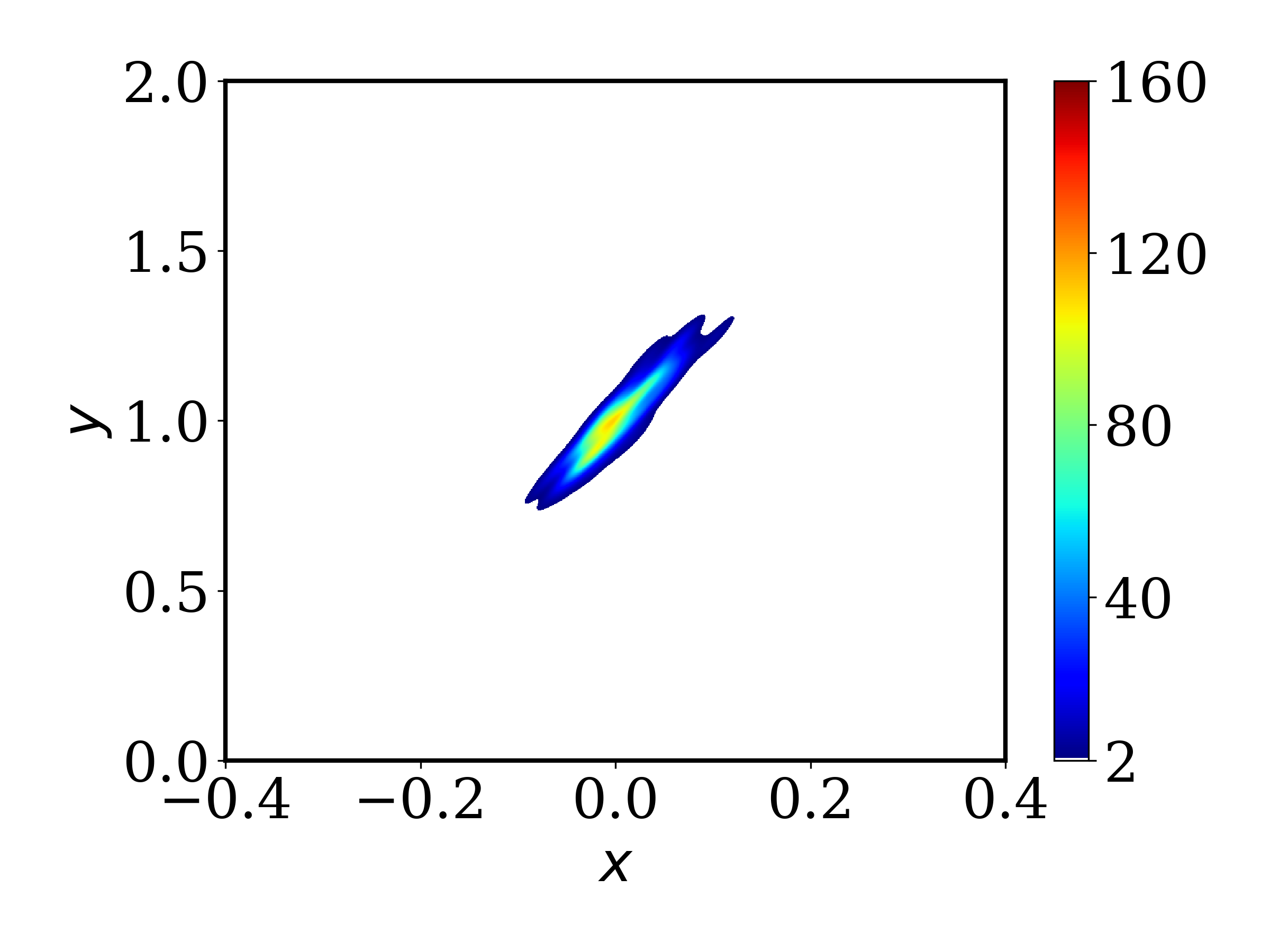}
        \caption{EnKF-MDA}
    \end{subfigure}
    \caption{Joint PDFs with $\bm{10^2}$ \textbf{samples} with the comparison among MCMC, EnKF, EnRML, and EnKF-MDA for the scalar case}
    \label{fig:scalar_limited_size}
\end{figure}

\begin{figure}[!htbp]
    \centering
    \includegraphics[width=0.8\textwidth]{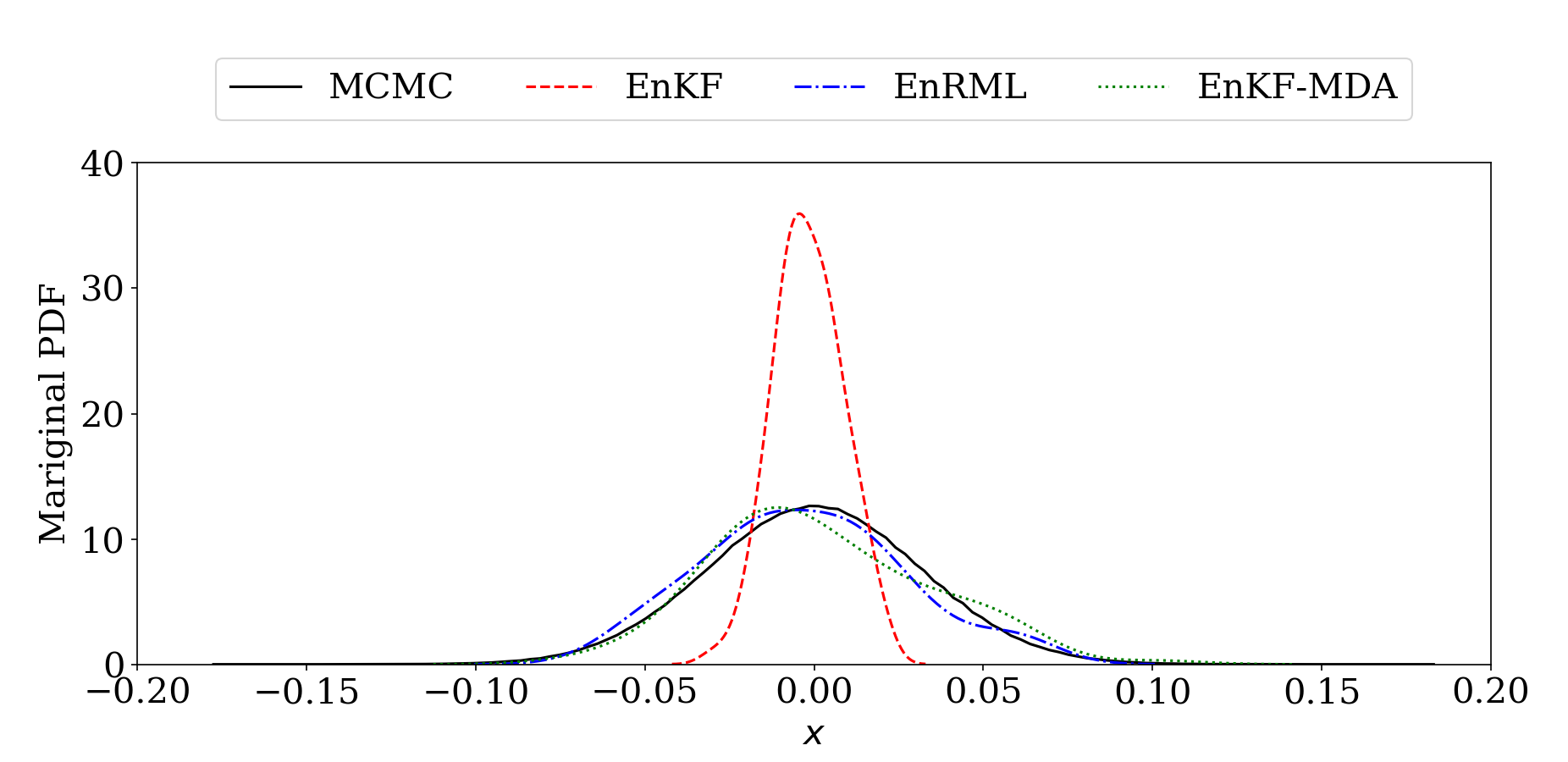}
    \caption{Marginal PDFs for $x$ with $\bm{10^2}$ \textbf{samples} with the comparison among MCMC, EnKF, EnRML, and EnKF-MDA for the scalar case.}
    \label{fig:marginal_pdf_smallsize}
\end{figure}

Further, we explore the effects of small ensemble size on this case and evaluate which method can outperform others with limited samples.
For many realistic cases, the propagation with large ensemble size is computationally prohibitive, and
ensemble methods can typically use less than $10^2$ samples to describe the statistical information. Therefore, we set the ensemble size to be $10^2$, and other set-ups are consistent with the previous case.
The joint PDF results with different ensemble methods are shown in Fig.~\ref{fig:scalar_limited_size}.
It can be seen that with the limited ensemble size, the iterative EnKF method performs similarly as with the large ensemble size. Specifically, all samples converge to the observations and the posterior distribution has a low variance.
By contrast, the EnRML method and EnKF-MDA not only can capture the posterior mean value but also provide the statistical information to indicate the uncertainty with ensemble realizations. 
For better visualization, the marginal PDFs in comparison of the three ensemble methods with $10^2$ samples are shown in Fig.~\ref{fig:marginal_pdf_smallsize}. We can see that the EnRML method and EnKF-MDA give satisfactory estimations on the uncertainty, while the mode value with EnKF is approximately three times higher than that with MCMC.
Generally, with limited ensemble size, EnKF performs similarly as with large ensemble size, which underestimates the posterior variance. The performance of EnRML and EnKF-MDA is still satisfactory but inferior to those with larger ensemble sizes.

\begin{figure}[!htbp]
    \centering
    \begin{subfigure}[b]{0.45\linewidth}
    \includegraphics[width=\textwidth]{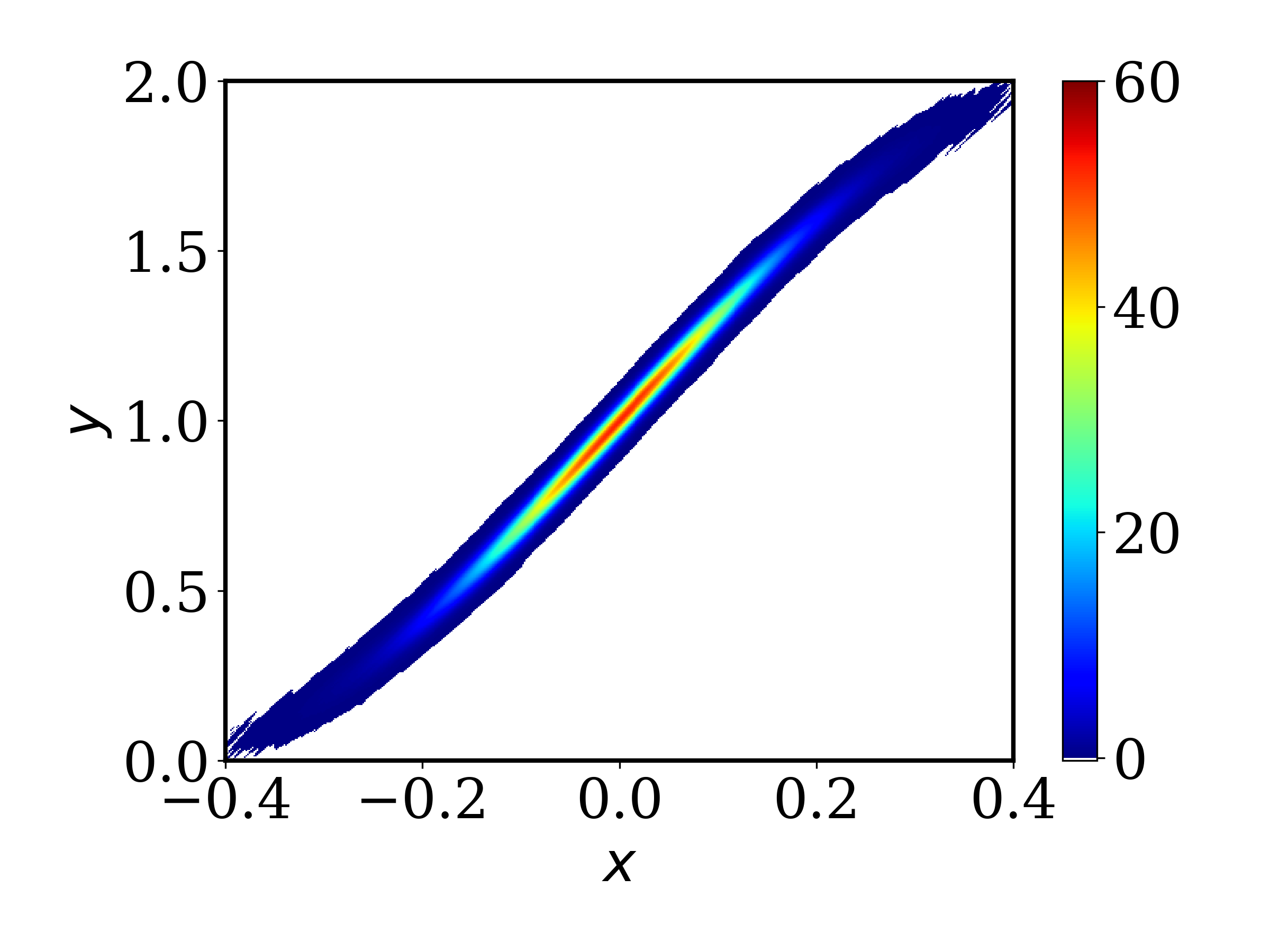}
    \caption{with $10^6$ samples}
    \end{subfigure}
    \begin{subfigure}[b]{0.45\linewidth}
    \includegraphics[width=\textwidth]{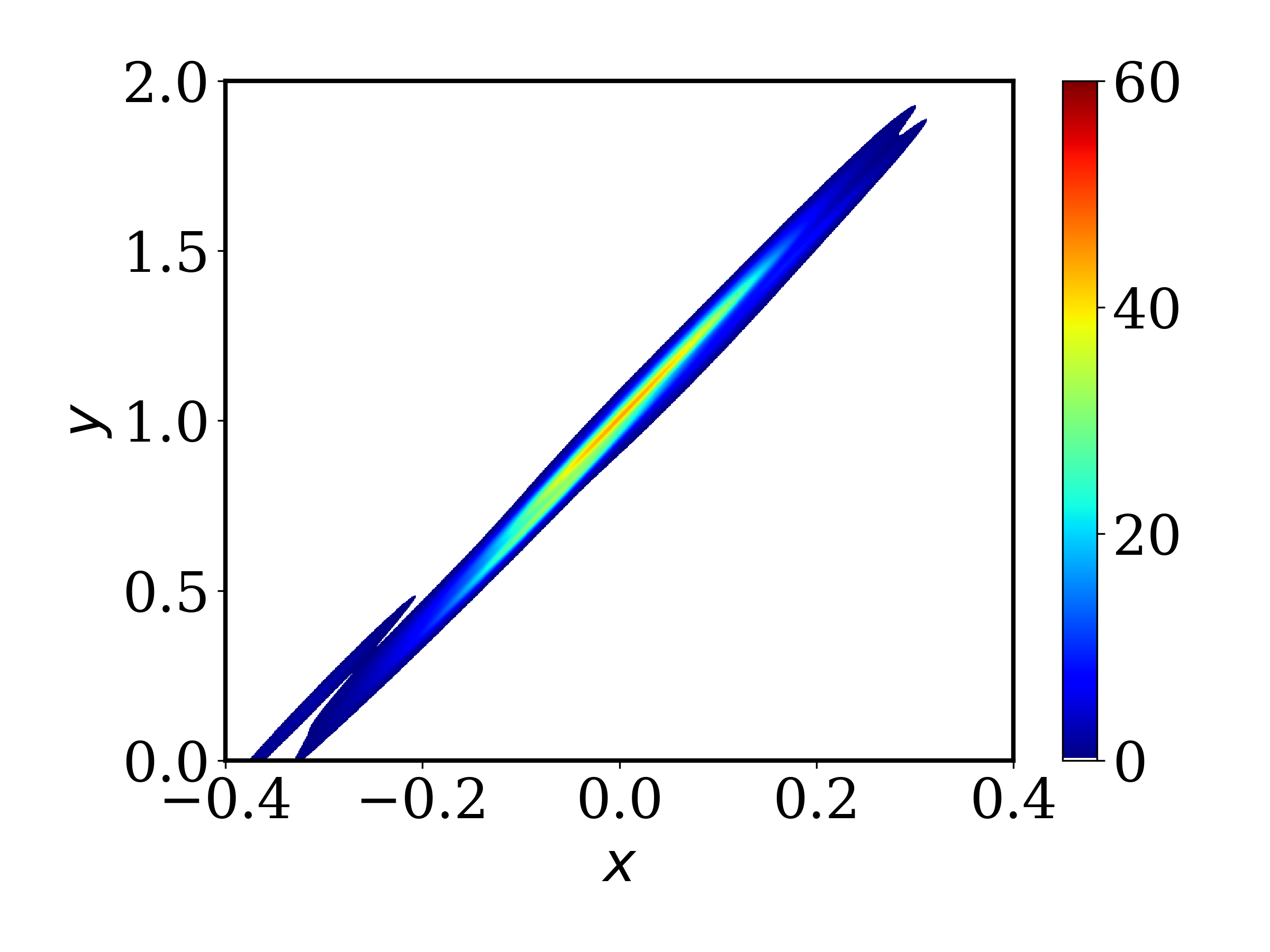}
    \caption{with $10^2$ samples}
    \end{subfigure}
    \caption{Results of prior joint PDF with large ($10^6$) and small ($10^2$) ensemble size for the scalar case}
    \label{fig:prior_compare}
\end{figure}

\begin{figure}[!htbp]
    \centering
    \begin{subfigure}[b]{0.45\linewidth}
    		\includegraphics[width=\textwidth]{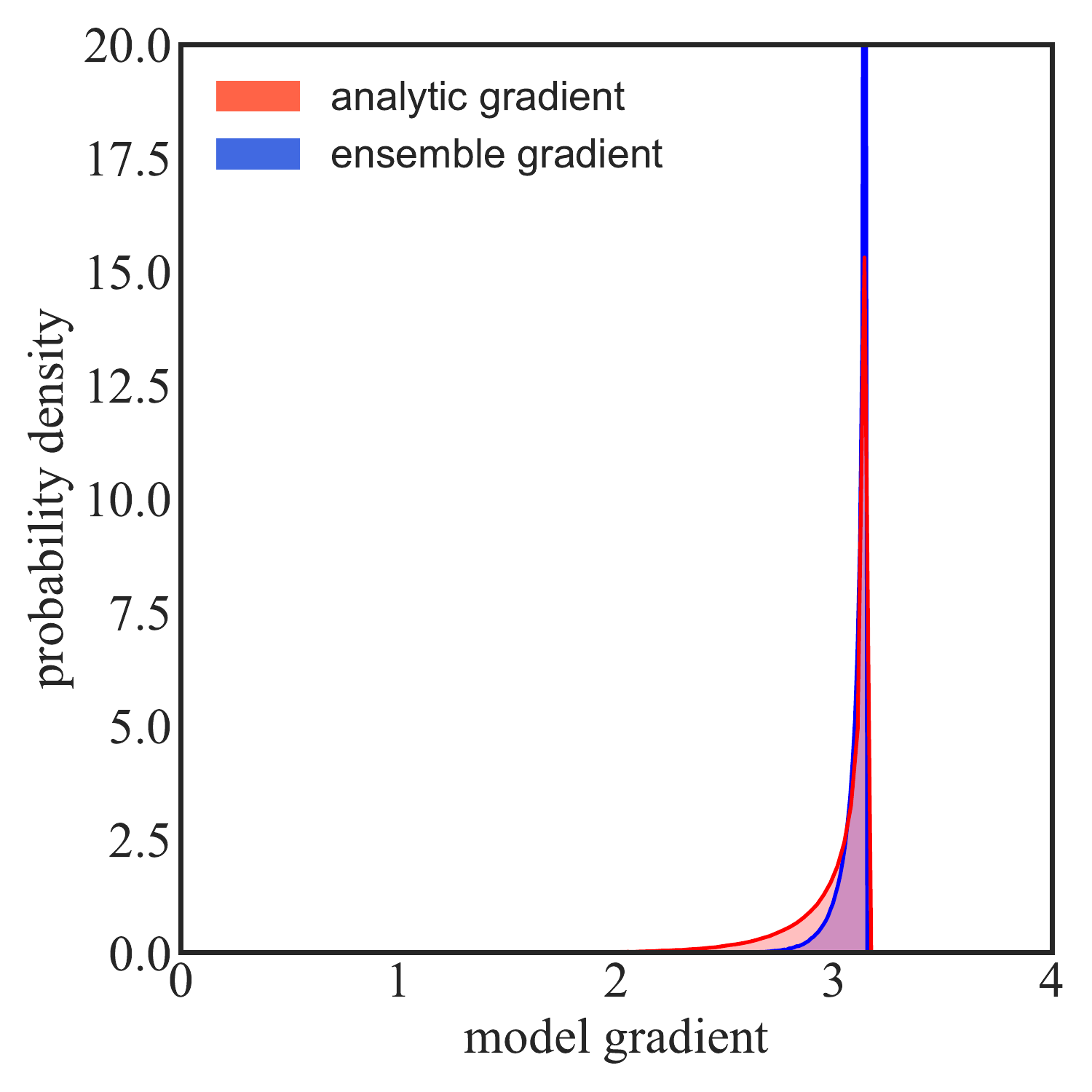}
    		\caption{with $10^6$ samples}
    \end{subfigure}
    \begin{subfigure}[b]{0.45\linewidth}
    		\includegraphics[width=\textwidth]{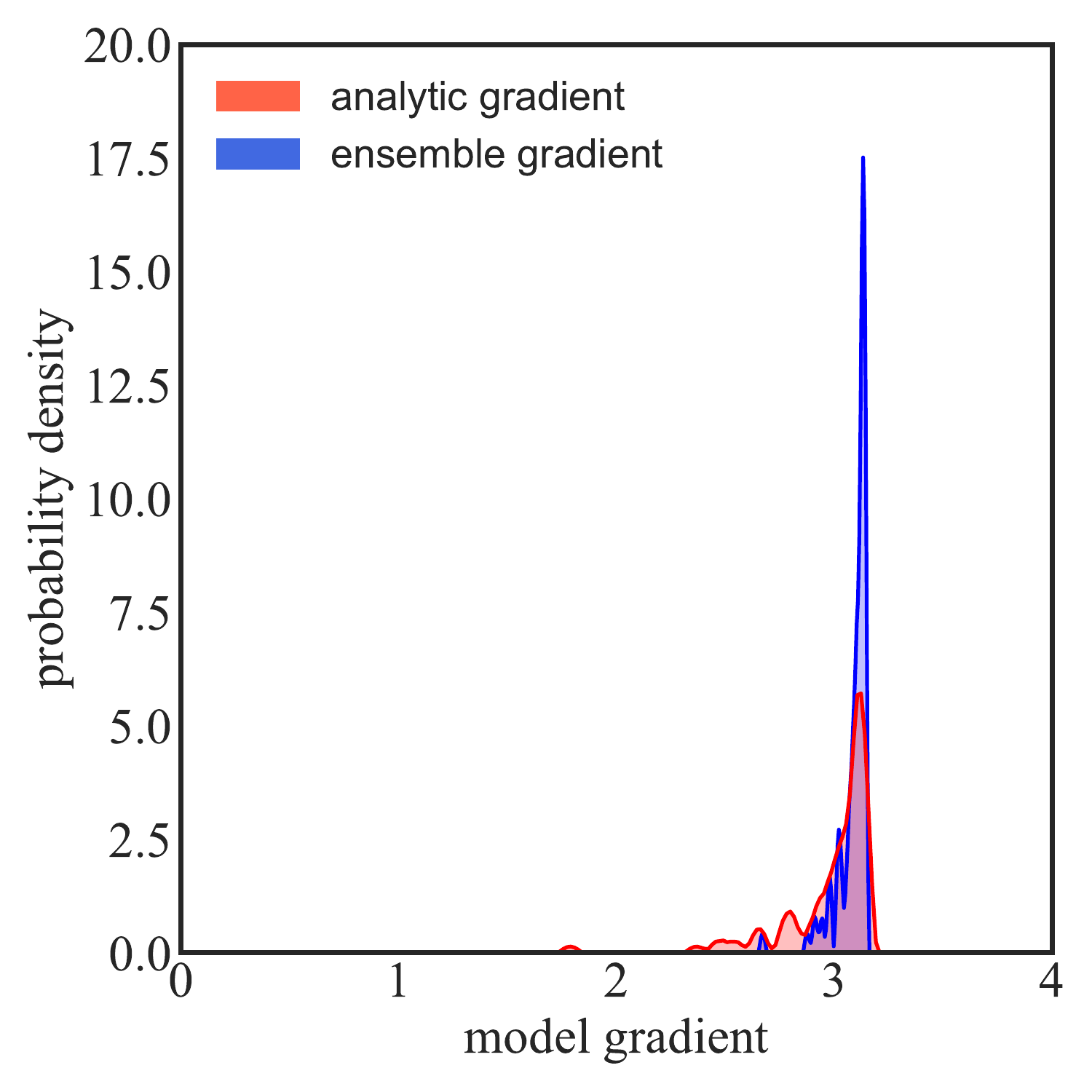}
    		\caption{with $10^2$ samples}
    	\end{subfigure}
    \caption{Comparison of analytic gradient and ensemble gradient. The light/pink shaded region represents analytic gradient and the dark/blue shaded region represents ensemble gradient. (a): $10^6$ samples; (b): $10^2$ samples}
    \label{fig:grad_compare}
\end{figure}

Not surprisingly, the estimation of uncertainty with limited ensemble size slightly deviates from the distribution obtained with MCMC. It is likely that the limited number of samples are insufficient for describing the necessary statistics. This may also increase the error in estimating the model gradient, especially for nonlinear models.
For illustration, we present the plots of prior joint PDF with the large and small ensemble size, as shown in Fig.~\ref{fig:prior_compare}. It is obvious that the small ensemble size is not sufficient to describe the prior distribution.
Additionally, we provide the model gradient estimated by ensemble samples in comparison with the analytic gradient.
The analytic gradient of this model is $\pi \cos\pi x$, and the ensemble gradient can be represented by $\frac{\sin(\pi \mathsf{x})-\sin(\pi \bar{\mathsf{x}})}{\mathsf{x}-\bar{\mathsf{x}}}$.
The sine function can be approximated as a linear model in the range close to zero, and thus we assume that:
\begin{align}
    \sin(\pi \mathsf{x})-\sin(\pi \bar{\mathsf{x}}) \approx \sin(\pi (\mathsf{x}-\bar{\mathsf{x}})),\quad \mathsf{x} \rightarrow 0 \text{,}
    \label{eq:scalar1}
\end{align}
and further
\begin{align}
     \lim_{\mathsf{x-\bar{\mathsf{x}} \to 0}} \frac{\sin(\pi (\mathsf{x}-\bar{\mathsf{x}}))}{\pi (\mathsf{x}-\bar{\mathsf{x}})} = \cos (\pi (\mathsf{x}-\bar{\mathsf{x}})).
     \label{eq:scalar2}
\end{align}
Based on this formula, we can see that if the samples are close to $\bar{\mathsf{x}}$ and the sample mean $\bar{\mathsf{x}}$ is estimated as zero, the ensemble gradient can be approximated to the analytic one as
\begin{equation*}
   \frac{\sin(\pi \mathsf{x})- \sin (\pi \bar{\mathsf{x}})}{\mathsf{x}-\bar{\mathsf{x}}} \stackrel{\eqref{eq:scalar1}}{\approx} \frac{\sin(\pi (\mathsf{x}-\bar{\mathsf{x}}))}{\mathsf{x}-\bar{\mathsf{x}}}
   \stackrel{\eqref{eq:scalar2}}{\approx}
   \pi  \cos(\pi (\mathsf{x}-\bar{\mathsf{x}}))
   \stackrel{\bar{\mathsf{x}} \approx 0}{\approx} \pi \cos(\pi \mathsf{x}).
\end{equation*}
Given that the model gradient is not subject to the Gaussian distribution, we use the cosine kernel to estimate the probability density, as shown in Fig.~\ref{fig:grad_compare}.
It is noticeable that the difference between the analytic gradient and ensemble gradient can be eased with the large ensemble size. The discontinuity in the case with $10^2$ samples is mainly due to the limited ensemble realizations which are insufficient to prescribe the infinite distribution.
The small ensemble size can significantly reduce the computational cost but may lead to additional errors in the statistical description and the model gradient estimation. To ensure the error remains within an acceptable range, the choice of the ensemble size need numerical tests. However, for highly nonlinear systems the reduction of errors in model gradient estimation will not benefit from large ensemble size unless the analytic gradient is adopted.
Also, localization techniques~\cite{anderson2012localization} can be introduced to reduce the sampling error and need future investigation.

\section{RANS equation}
\label{sec:RANS}
CFD is of significant importance for many engineering applications to inform the process of design, analysis, and optimization.
Considering the computational cost, the RANS model is still the primary tool to characterize turbulence behavior in CFD simulations.
However, the unknown Reynolds stress term in RANS equations is commonly solved with different closure models under the Boussinesq assumption. This assumption introduces the model uncertainty and reduces the confidence on the predictive performance.
In this section, we apply the three ensemble-based data assimilation methods (EnKF, EnRML, and EnKF-MDA) on the RANS closure problem and evaluate their performance to quantify and reduce the uncertainty of the predicted velocity by incorporating high fidelity data.

\subsection{Problem statement}
The RANS equations can be expressed as:
\begin{subequations}
\label{eq:rans}
\begin{align}
\quad \frac{\partial U_i}{\partial x_i} & = 0  \\
\frac{\partial U_i}{\partial t}+\frac{\partial \left(U_i U_j \right)}{\partial x_j}
                                            & =  -\frac{\partial {P}}{\partial x_i}
                                              +\frac{1}{Re}\frac{\partial^2 U_i}{\partial x_j \partial  x_j}
                                              - \frac{\partial \overline{  u_i'  u_j'}}{\partial x_j}  ,
\label{eq:rans-momentum}
\end{align}
\end{subequations}
where~$U, P$ is the dimensionless velocity and pressure respectively, and $Re$ is the Reynolds number. 
In the momentum equation~\eqref{eq:rans-momentum},~$\tau = -\overline{u_i' u_j'}$ is the Reynolds stress which is the main source of uncertainty in RANS simulations. 
We regard the Reynolds stress from RANS simulation coupling with the linear eddy--viscosity model as the baseline. 
Further, we introduce the discrepancy term~$\Delta \tau$ representing the uncertainty into the baseline as
\begin{equation}
     \tau \equiv \tau^{\text{RANS}} + \Delta \tau.
     \label{eq:Reynolds_stress_disc}
\end{equation}
Thus, we can quantify the uncertainty in the predicted velocity with the three ensemble-based DA methods by incorporating available observation data.

\subsection{Methodology}
The data assimilation framework to quantify and reduce the RANS model-form uncertainty associated with Reynolds stress was proposed by Xiao et.al~\cite{xiao_quantifying_2016}. 
Here, we give a brief introduction to this methodology, and the reader is referred to~\cite{xiao_quantifying_2016} for further details.

\begin{figure}
    \centering
    \begin{subfigure}[b]{0.45\linewidth}
    	\includegraphics[width=\textwidth]{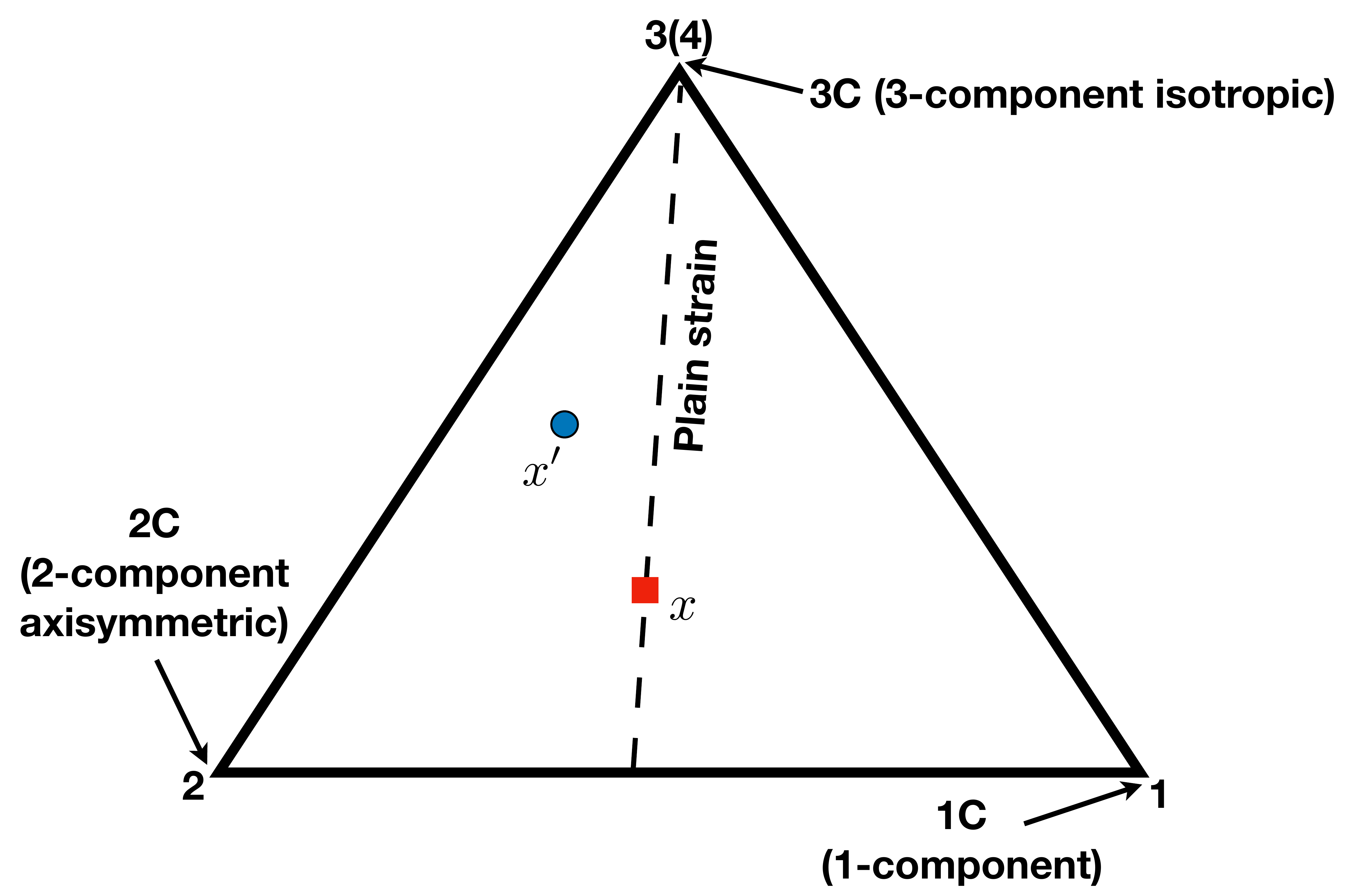}
    	\caption{Barycentric coordinate}
    	\label{fig:barycentric_triangle}
    \end{subfigure}
    \begin{subfigure}[b]{0.35\linewidth}
    	\includegraphics[width=\textwidth]{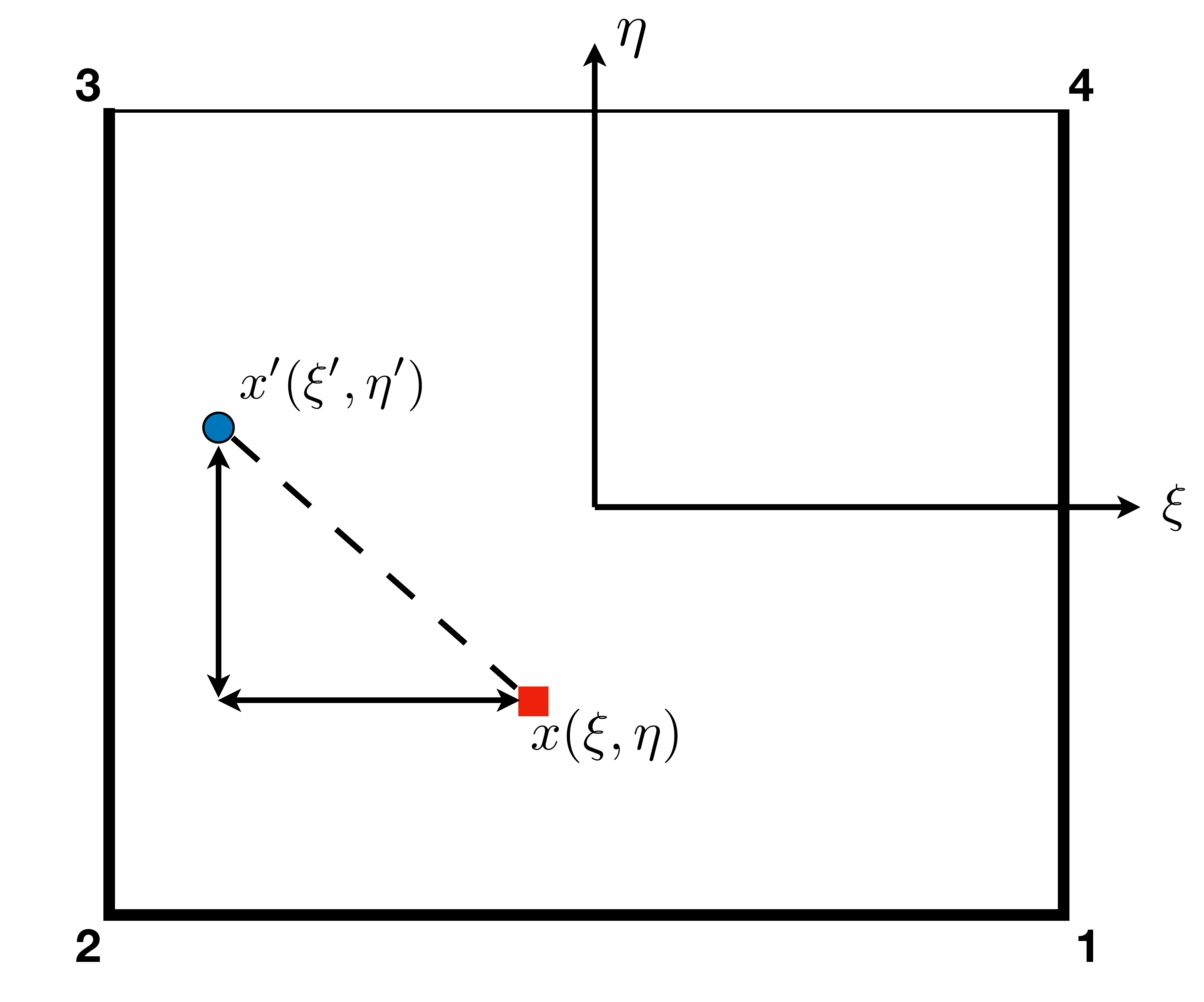}
    	\caption{Natural coordinate}
    	\label{fig:natural_coor}
    \end{subfigure}    
    \caption{Mapping between the barycentric coordinate to the natural coordinate}
\end{figure}

To quantify the uncertainty within Reynolds stress, we first transform the Reynolds stress tensor into several scalar fields.
Specifically, the Reynolds stress tensor can be expressed as
\begin{equation*}
\tau = 2k(\frac{1}{3} \textbf{I} + \textbf{a})=2k (\frac{1}{3} \textbf{I} + \textbf{V} \Lambda \textbf{V}^\top) \text{,}
\end{equation*}
where $k$ is the turbulent kinetic energy, indicating the magnitude of the Reynolds stress, $\textbf{I}$ is the second order identity tensor, $\textbf{a}$ is the anisotropy tensor; $\textbf{V}=[\textbf{v}_1, \textbf{v}_2, \textbf{v}_3]$, and $\Lambda = \text{diag}[\lambda_1, \lambda_2, \lambda_3]$ with $\lambda_1 + \lambda_2 + \lambda_3 = 0$ are the eigenvector and eigenvalue of $\textbf{a}$, respectively, which represents the shape and orientation of $\tau$.
Afterwards, the eigenvalues $\lambda_1, \lambda_2, \lambda_3$ are projected to a barycentric coordinate as
\begin{align*}
	C_1 &= \lambda_1 - \lambda_2 \\
	C_2 &= 2(\lambda_2 - \lambda_3) \\
	C_3 &= 3 \lambda_3 + 1 ,
\end{align*}
with $C_1 + C_2 + C_3=1$.~\cite{banerjee2007presentation}
The barycentric coordinate is shown in Fig.~\ref{fig:barycentric_triangle}.
To facilitate the parameterization, the barycentric coordinate is transformed to the natural coordinate $\bm{\chi} = (\xi, \eta)$ by placing the triangle in a Cartesian coordinate as shown in Fig.~\ref{fig:natural_coor}.
The location of any point in the triangle can be expressed as a combination of those of the three vertices. That is,
\begin{equation}
	\bm{\chi}=\bm{\chi}_{1c} C_1 + \bm{\chi}_{2c}C_2 + \bm{\chi}_{3c}C_3 ,
\end{equation}
where $\bm{\chi}_{1c}$, $\bm{\chi}_{2c}$, and $\bm{\chi}_{3c}$ are the coordinates of the three vertices of the triangle.

In conclude, we represent the Reynolds stress baseline~$\tau^\text{RANS}$ with three discrepancy variables~$k^\text{RANS}$,~$\xi^\text{RANS}$, and~$\eta^\text{RANS}$ through eigendecomposition and coordinate conversion.
Further, the additive uncertainties~$\delta^k$,~$\delta^\epsilon$, and~$\delta^\eta$ can be injected into these projected variables as
\begin{subequations}
	\begin{align}
	\log k(x)&= \log k^{\text{RANS}}(x) + \delta^k(x),\\
	\xi(x)&= \xi^{\text{RANS}}(x) + \delta^\xi(x),\\
	\eta(x)&= \eta^{\text{RANS}}(x) + \delta^\eta(x) ,
	\end{align}
\end{subequations}
where the logarithm on~$k$ is to ensure non-negativity. 
The dimension of the variables~$\log k(x)$,~$\xi(x)$, and~$\eta(x)$ is consistent with the mesh grid.
To infer the entire field with very sparse observation significantly increases the ill-posedness of the problem. 
Hence, it is necessary to reduce the dimension of the state space. 
In this case, we leverage the Karhunen–-Lo\`eve (KL) expansion with truncated orthogonal modes to represent the field for each quantity to be inferred. 
Concretely, the discrepancy variables~$\delta^k$,~$\delta^\xi$, and~$\delta^\eta$ are constructed as the random fields subject to zero-mean Gaussian process~$\mathcal{GP}(0, \mathcal{K})$. 
The kernel function~$\mathcal{K}$ indicates the covariance at two locations $x$ and $x'$ as
\begin{equation}
    \mathcal{K}(x,x') = \sigma(x)\sigma(x')\exp(-\frac{|x-x'|^2}{l^2}).
\end{equation}
In the formula above,~$\sigma(x)$ is the variance field to reflect the region where large discrepancy is expected.~$l$ is the characteristic length.
The KL modes take the form as:~$\phi_i(x)=\sqrt{\hat{\lambda}}\hat{\phi}_i(x)$, where~$\hat{\lambda}$ and~$\hat{\phi}$ are the eigenvalues and eigenvectors of the kernel~$\mathcal{K}$, respectively,
computed based on the Fredholm integral as
\begin{equation}
    \int \mathcal{K}(x,x')\hat{\phi}(x')dx' = \hat{\lambda}\hat{\phi}(x) \text{.}
\end{equation}
This choice of KL modes for the discrepancy fields $\delta^k$, $\delta^\xi$, $\delta^\eta$ leads to a KL expansion.
That is, the discrepancy variables can be constructed from these deterministic KL modes~$\phi(x)$ and zero-mean, uni-variance random variable~$\omega$ as
\begin{equation}
\begin{aligned}
     \delta^k(x) &= \sum_{i=1}^N \omega^k_i\phi_i(x),\\
     \delta^\xi(x) &= \sum_{i=1}^N \omega^\xi_i\phi_i(x),\\
     \delta^\eta(x) &= \sum_{i=1}^N \omega^\eta_i\phi_i(x).
\end{aligned}
\label{eq:delta}
\end{equation}
 With~$\omega$ and KL modes~$\phi(x)$, we can reconstruct the field of each discrepancy quantity and recover the random field of Reynolds stress tensor.

The Reynolds stress representation and dimension reduction presented above makes it practical to quantify and reduce the uncertainty in the RANS model by incorporating observation data, i.e., direct numerical simulation (DNS) results. 
From a Bayesian perspective, the random noise $\eta \sim (0, \sigma_\text{obs} )$ is added in time-averaged DNS data $\mathsf{y}$ to allow overlap between the likelihood and the prior distribution.
Herein the $\sigma_\text{obs}$ is the standard deviation of observation noise, indicating the noise level.
We take the velocity as the state augmented with the KL coefficients.
As a result, we can adopt the iterative ensemble methods (EnKF, EnRML, and EnKF-MDA) to quantify and reduce the uncertainty in velocity with prior samples of the KL coefficient and the observation.

In summary, the procedure of the RANS model-form uncertainty quantification framework is presented below:
\begin{enumerate}
\item \textbf{Preprocessing step:}\\
(1) Perform RANS simulation to obtain~$\tau^\text{RANS}$ as the baseline.\\
(2) Project~$\tau^\text{RANS}$ onto the field of~$k^\text{RANS}$,~$\xi^\text{RANS}$, and~$\eta^\text{RANS}$.\\
(3) Conduct KL expansion to generate the KL basis sets or modes~$\{\phi_i(x)\}_{i=1}^m$, where~$m$ is the number of truncated modes.\\
(4) Generate the initial value of~$\omega$ with a zero-mean uni-variance normal distribution.
\item \textbf{Data assimilation step:}\\
(a) Recover the discrepancy fields of~$\delta^k$,~$\delta^\xi$, and~$\delta^\eta$ with coefficient~$\omega$ and basis sets~$\phi(x)$ based on Eq.~\eqref{eq:delta}.\\
(b) Reconstruct the ensemble realizations on~$\tau$ through mapping~$(k, \xi, \eta) \rightarrow \tau$ and solve the RANS equation to obtain the velocity field given each realization of~$\tau$. \\
(c) Perform the Bayesian analysis with data assimilation technique to reduce the uncertainty of velocity by incorporating time-averaged DNS data.\\
(d) Return to step~(a) till the convergence criteria or maximum iteration number is reached.
\end{enumerate}

\subsection{Case setup}
\begin{figure}
    \centering
    \includegraphics[width=0.6\textwidth]{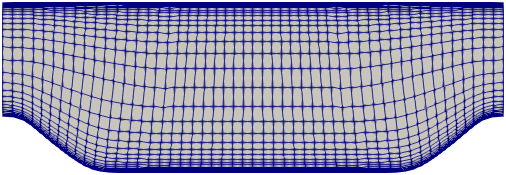}
    \caption{The structured mesh used for the simulation of flow over the periodic hills}
    \label{fig:pehill_mesh}
\end{figure}

The test case is turbulent flow over periodic hills initially proposed by Fröhlich et al.~\cite{frohlich2005highly}. 
The Reynolds number based on the bulk velocity and height of crest is~$2800$.
We use the DNS data from~\cite{breuer2009flow} as the benchmark.
The Launder--Sharma RANS model~\cite{launder1974application} is one of classical low Reynolds $k$--$\epsilon$ models and is extensively used in industrial applications.
Hence, we use the RANS simulation with this model as the baseline.
The periodic boundary condition is imposed on the inlet, and the non-slip boundary condition is applied on the wall.
A structured mesh is constructed with~$50$ cells in the stream-wise direction and~$30$ cells in the normal to wall direction, as shown in Fig.~\ref{fig:pehill_mesh}. Despite the coarse mesh, the dimensionless distance $y^+$ between the first cell and the walls is around $1$, which meets the requirement of the Launder--Sharma turbulence model.

As for the data assimilation setup, the number of KL modes for~$k$,~$\xi$, and~$\eta$ is set to~$8$. 
The ensemble size is~$50$.
The length scale is set as constant~$1$ for simplicity.
The standard deviation of observation noise $\sigma_\text{obs}$ is set as~$10\%$ of the truth.
We take~$18$ observations to quantify and reduce the uncertainty in velocity. 
The locations are marked in Fig.~\ref{fig:CFD_Prior}.
The step parameter~$\gamma$ in the EnRML method is chosen as $0.5$, and the inflation parameter~$N_{\text{mda}}$ in EnKF-MDA is set as~$50$ to obtain the convergence results based on our calibration study.
For this case, the MCMC sampling is impractical to verify the estimated posterior uncertainty, due to the high dimensionality of the state space and the high costs of numerical simulation.

The built-in solver \textit{simpleFoam} in OpenFoam is used to run the RANS simulation and obtain the baseline/prior mean.
The forward solver \textit{tauFoam} is developed based on \textit{simpleFoam} to propagate the Reynolds stress to velocity. That is, the forward solver computes velocity with the given Reynolds stress field rather than using turbulence models.

\subsection{Results}

\begin{figure}[!htbp]
    \centering
    \includegraphics[width=0.5\textwidth]{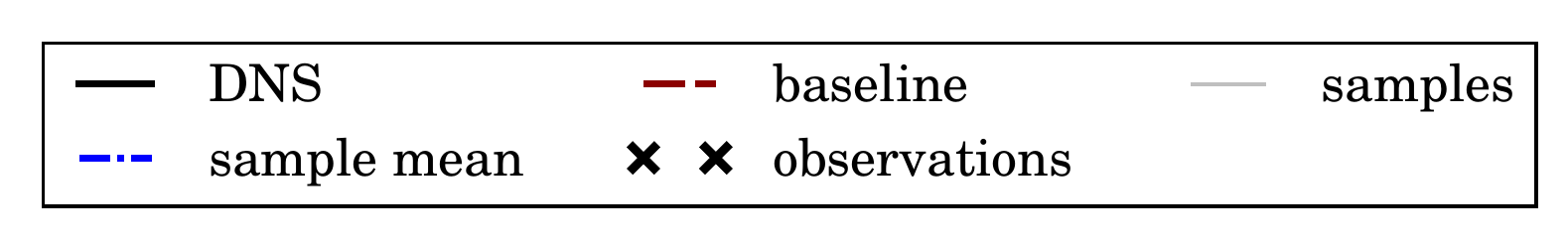}\\
    \includegraphics[width=0.7\textwidth]{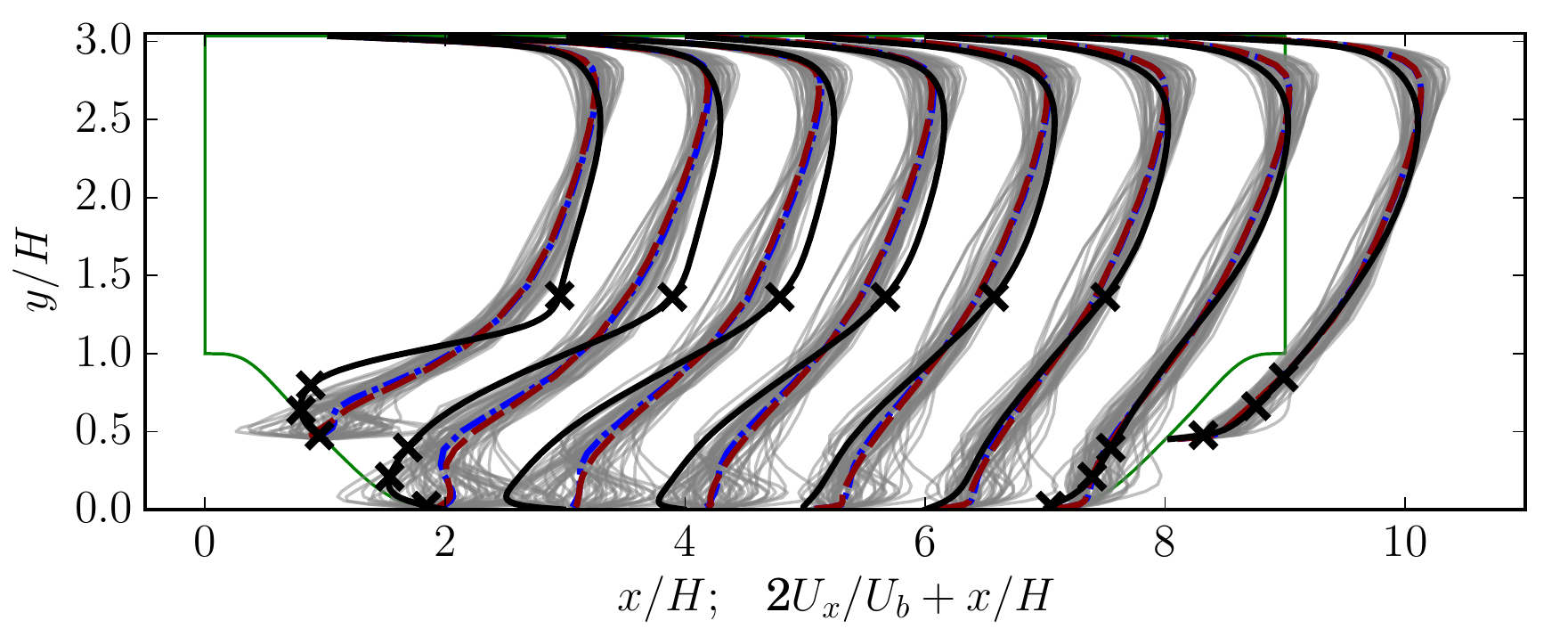}
    \caption{Prior ensemble realization of stream-wise velocity profiles at~$18$ locations, in comparison to DNS and baseline. The location of observation is indicated with crosses($\times$).}
    \label{fig:CFD_Prior}
\end{figure}

Through solving RANS equations given the randomized Reynolds stresses, we can obtain the prior uncertainty in the propagated velocity.
The plots of the prior stream-wise velocity are shown in Fig.~\ref{fig:CFD_Prior}.
It can be seen that the space spanned by the ensemble realizations can indicate the statistical information.
Also, the sample mean can have a good fit with RANS results.
That is reasonable since the random field is constructed by perturbing the baseline from RANS simulation.

\begin{figure}[!htbp]
    \centering
    \includegraphics[width=0.5\textwidth]{figure11-legend.pdf}\\
    \begin{subfigure}[b]{0.6\linewidth}
    		\includegraphics[width=\textwidth]{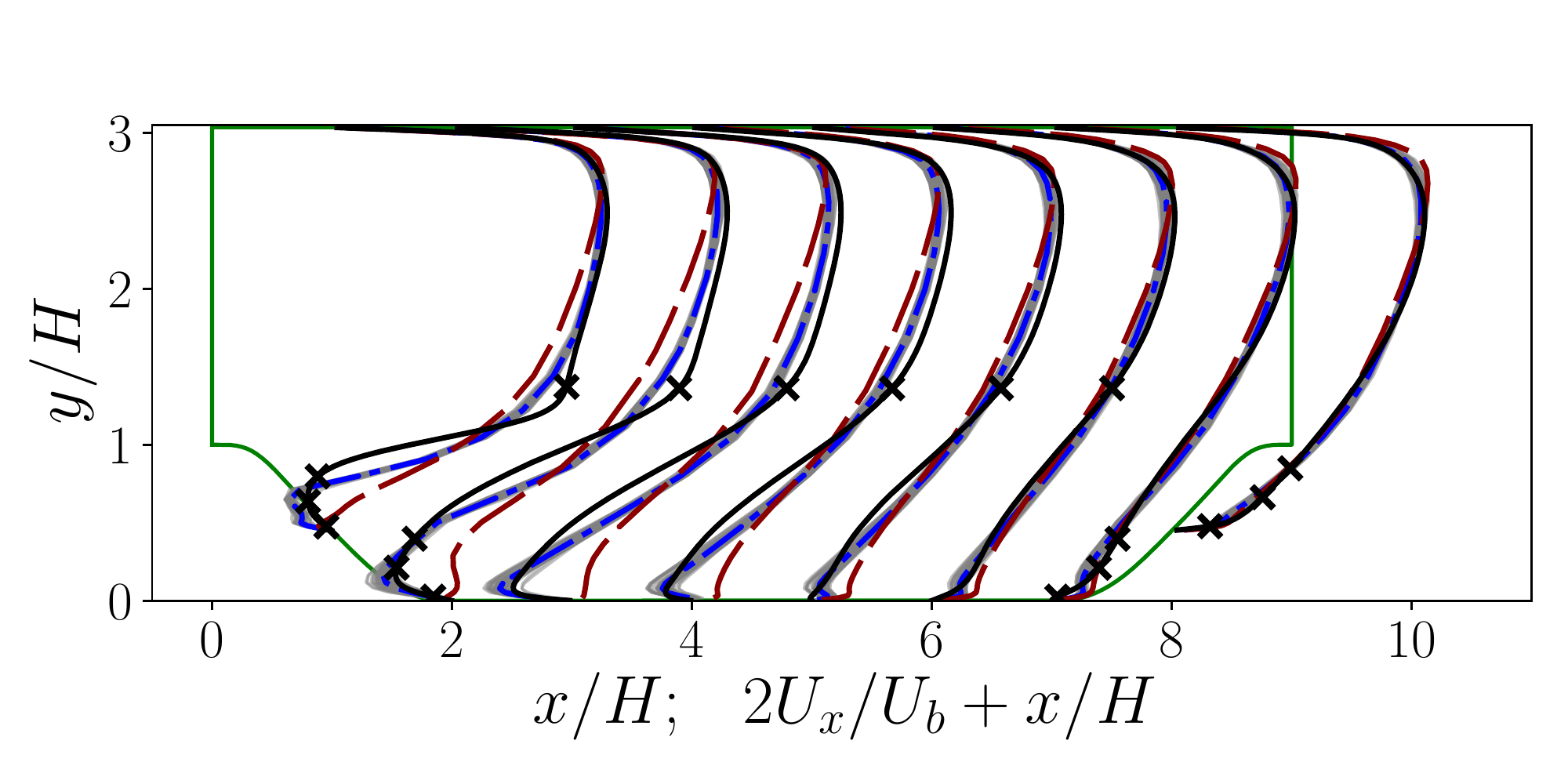}
    		\caption{EnKF}
    \end{subfigure}
    \begin{subfigure}[b]{0.6\linewidth}
    		\includegraphics[width=\textwidth]{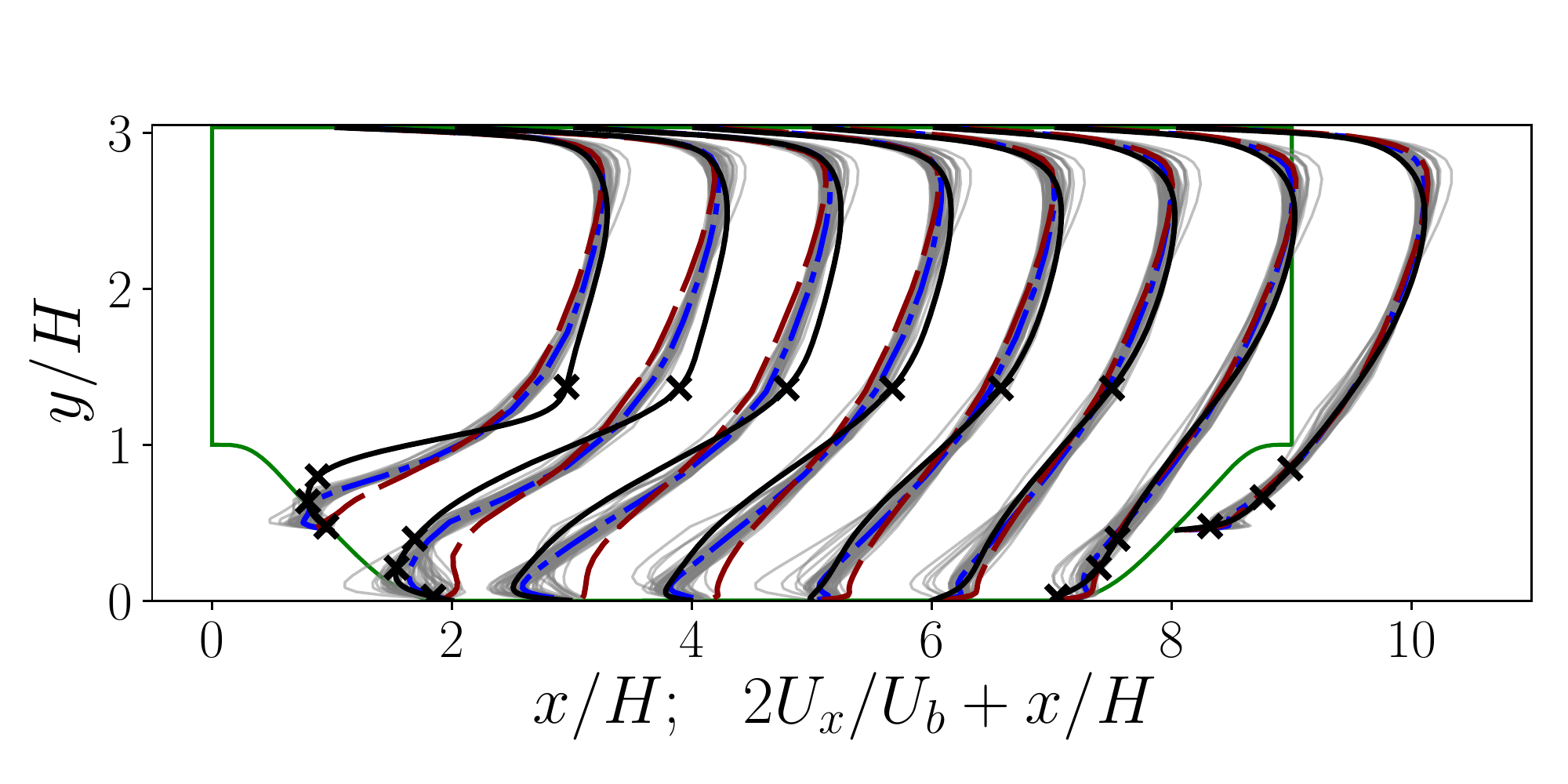}
    		\caption{EnRML}
    \end{subfigure}
    \begin{subfigure}[b]{0.6\linewidth}
    		\includegraphics[width=\textwidth]{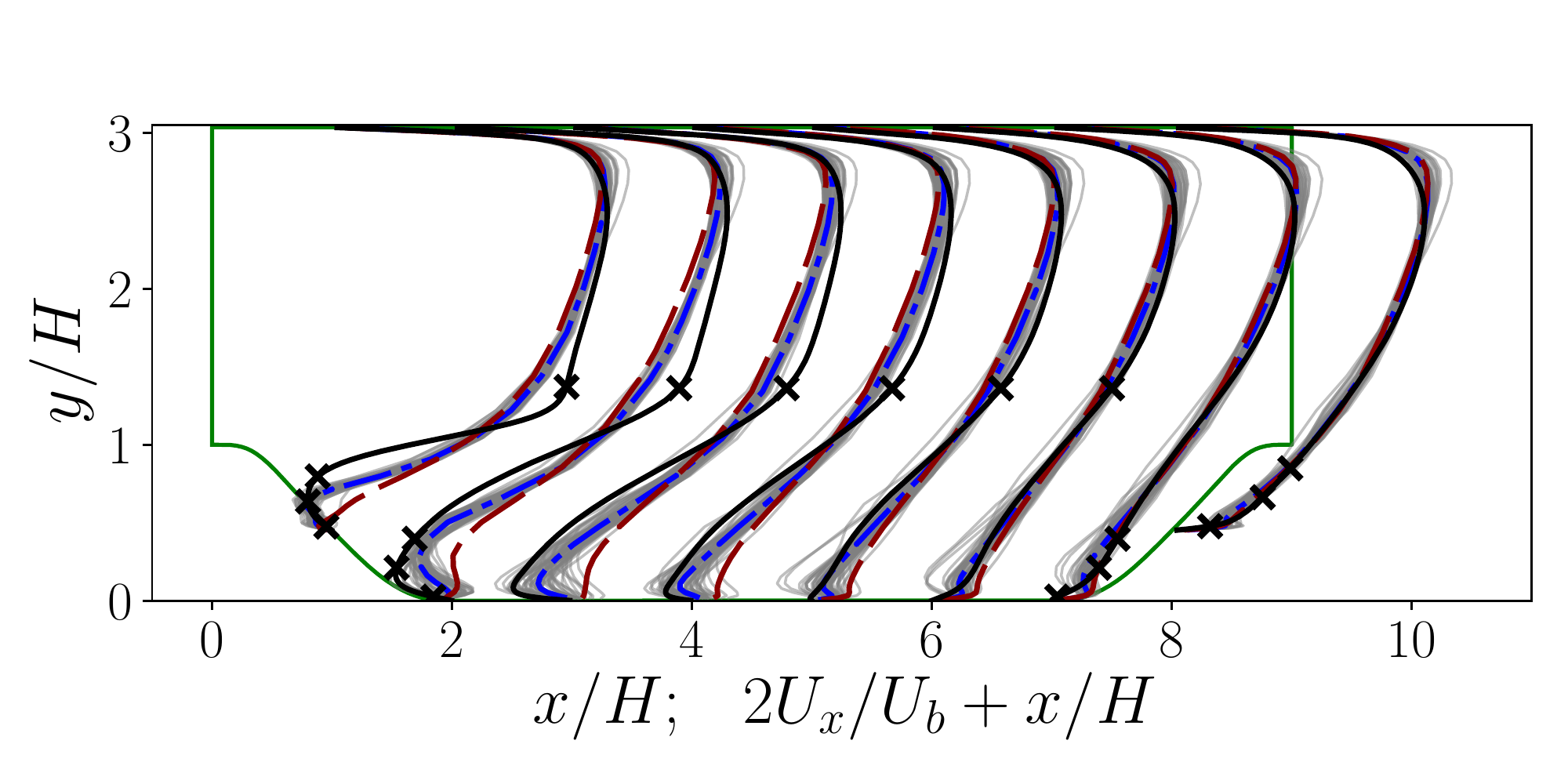}
    		\caption{EnKF-MDA}
    \end{subfigure}
    \caption{Data assimilation results of stream-wise velocity  with EnKF, EnRML, and EnKF-MDA in comparison to baseline and DNS for the turbulent flow in a periodic hill.}
    \label{fig:CFD_case_posterior}
\end{figure}

Further, we perform data assimilation analysis with EnKF, the EnRML method, and EnKF-MDA to quantify uncertainties in the velocity field by incorporating the observations at the specific locations.
The results with different data assimilation schemes are presented in Fig.~\ref{fig:CFD_case_posterior}.
It is noticeable that with EnKF the posterior mean can fit well with DNS results. However, all samples converge to the mean value, and the variance of the posterior becomes very low. By contrast, the EnRML method can give an estimation of the uncertainty, and the mean value also has a good fit with DNS data.
EnKF-MDA can also preserve the sample variance and improve the data fit, but the sample mean is relatively inferior compared to the other two methods.
Based on our derivation and evaluation in the former sections, that is likely due to EnKF repeatedly using the same DNS data with full Gauss--Newton steps, while the EnRML method and EnKF-MDA can be considered to perform one EnKF step via several small analysis steps.

\begin{figure}[!htbp]
    \centering
    \includegraphics[width=0.4\textwidth]{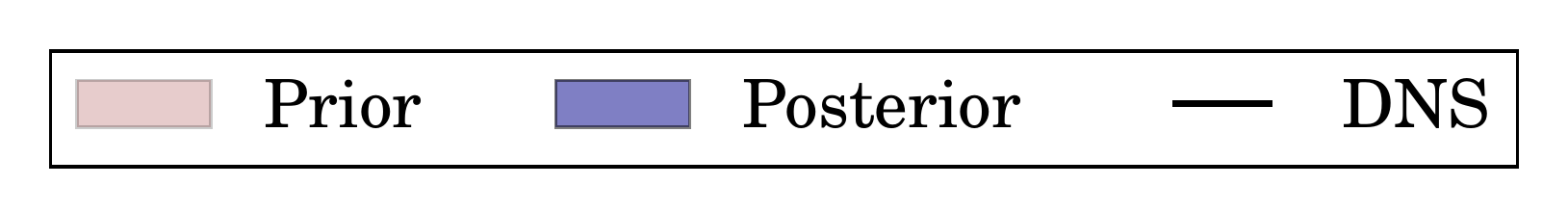}\\
    \begin{subfigure}[b]{0.6\linewidth}
    		\includegraphics[width=\textwidth]{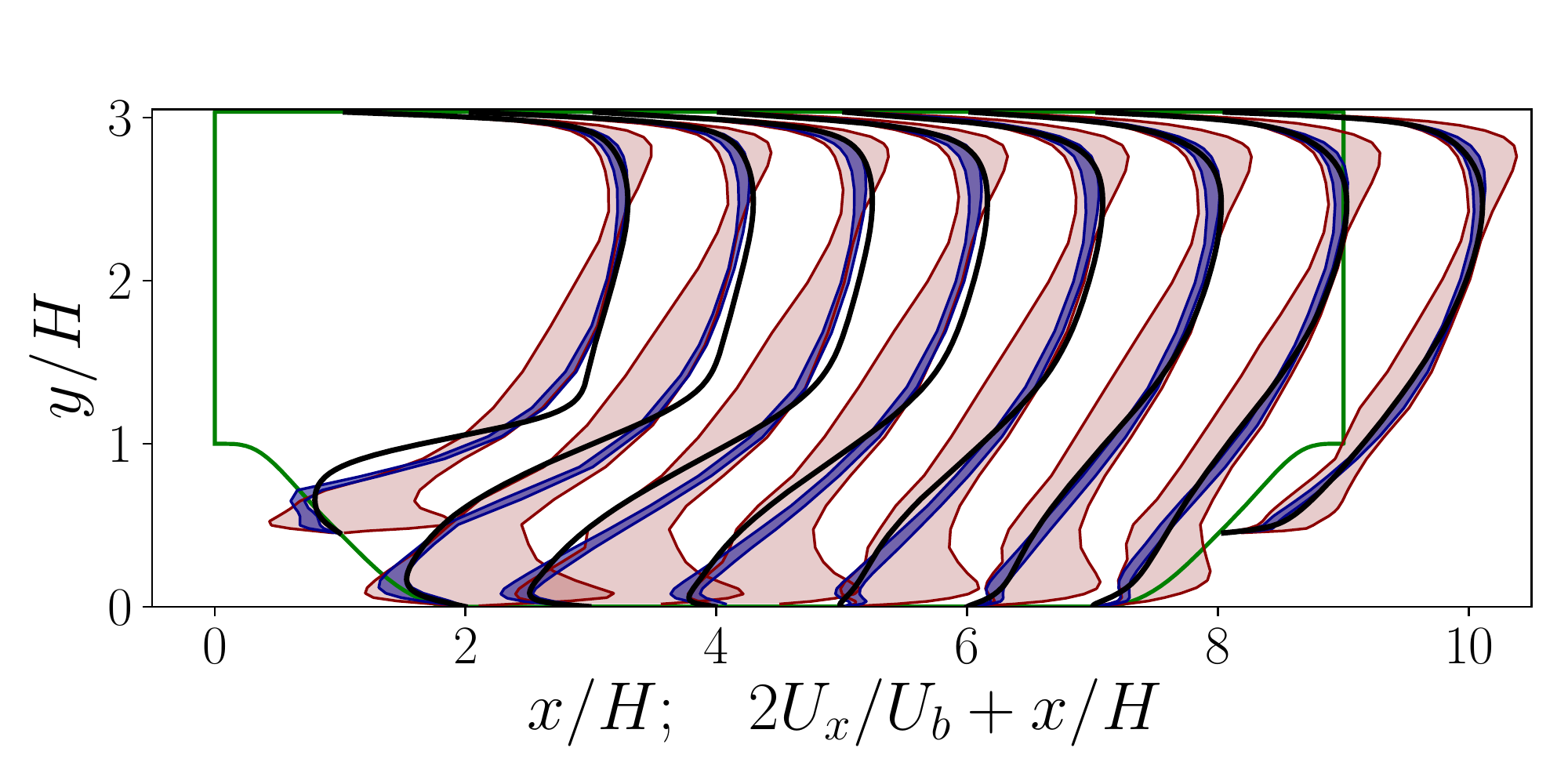}
    		\caption{EnKF}
    \end{subfigure}
    \begin{subfigure}[b]{0.6\linewidth}
    	 	\includegraphics[width=\textwidth]{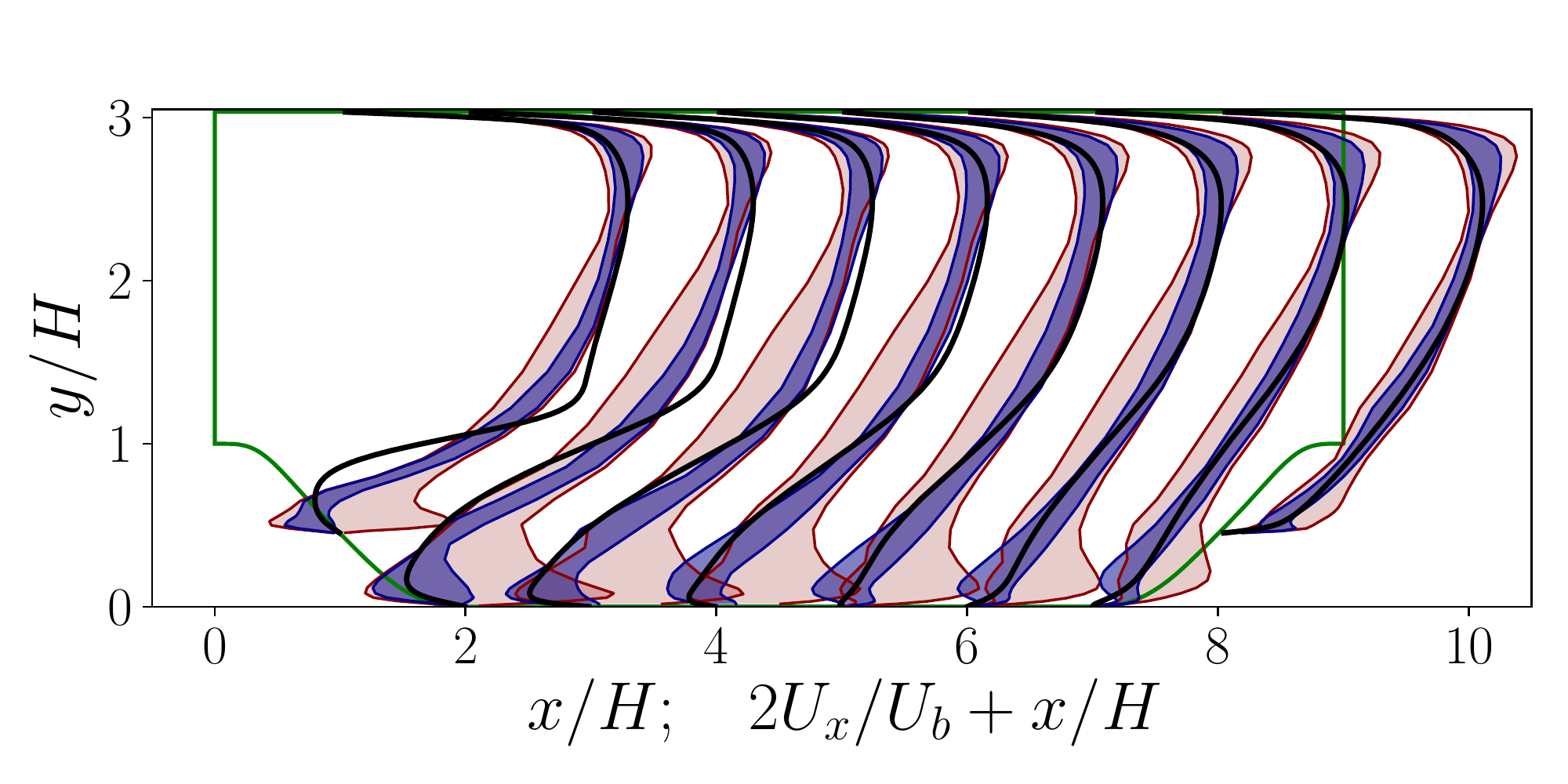}
    		\caption{EnRML}
    \end{subfigure}
    \begin{subfigure}[b]{0.6\linewidth}
    		\includegraphics[width=\textwidth]{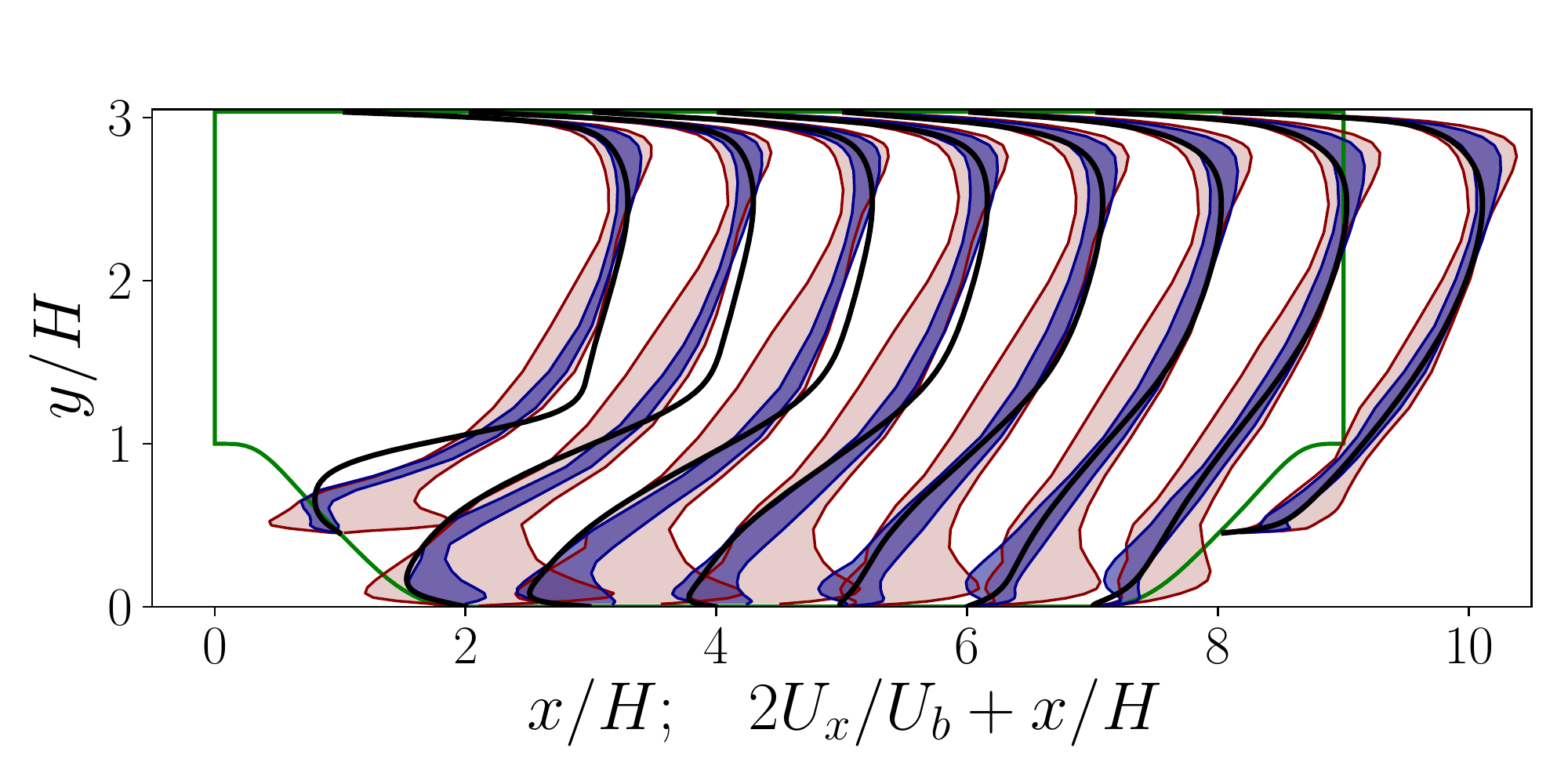}
    		\caption{EnKF-MDA}
    \end{subfigure}
    \caption{The $95\%$ credible intervals of the prior (light/pink shaded region) and posterior (dark/blue shaded region) samples of stream-wisevelocity profiles for the turbulent flow in a periodic hill}
    \label{fig:CFD_confidence}
\end{figure}

Here we present the comparison of~$95\%$ credible interval between the prior and posterior with the three data assimilation methods.
The results are shown in Fig.~\ref{fig:CFD_confidence}. It is noticeable that the posterior uncertainty with EnKF is underestimated and too much confidence is placed in the mean value. With the EnRML method and EnKF-MDA, we can have an estimation of the uncertainty indicated by samples. 
Besides, the uncertainty in the upper channel estimated by the EnRML method and EnKF-MDA is similar to the prior. 
That is reasonable since the variance~$\sigma$ in this region is low~\cite{xiao_quantifying_2016}, and no observation is informed as well. Hence, the posterior should not change much from the prior distribution.
Based on the overall performance, the iterative EnKF loss the statistical information due to data overuse, while the other two methods can provide reasonable uncertainty information. 

\begin{figure}
    \centering
    \begin{subfigure}[b]{0.45\linewidth}
    	 	\includegraphics[width=\textwidth]{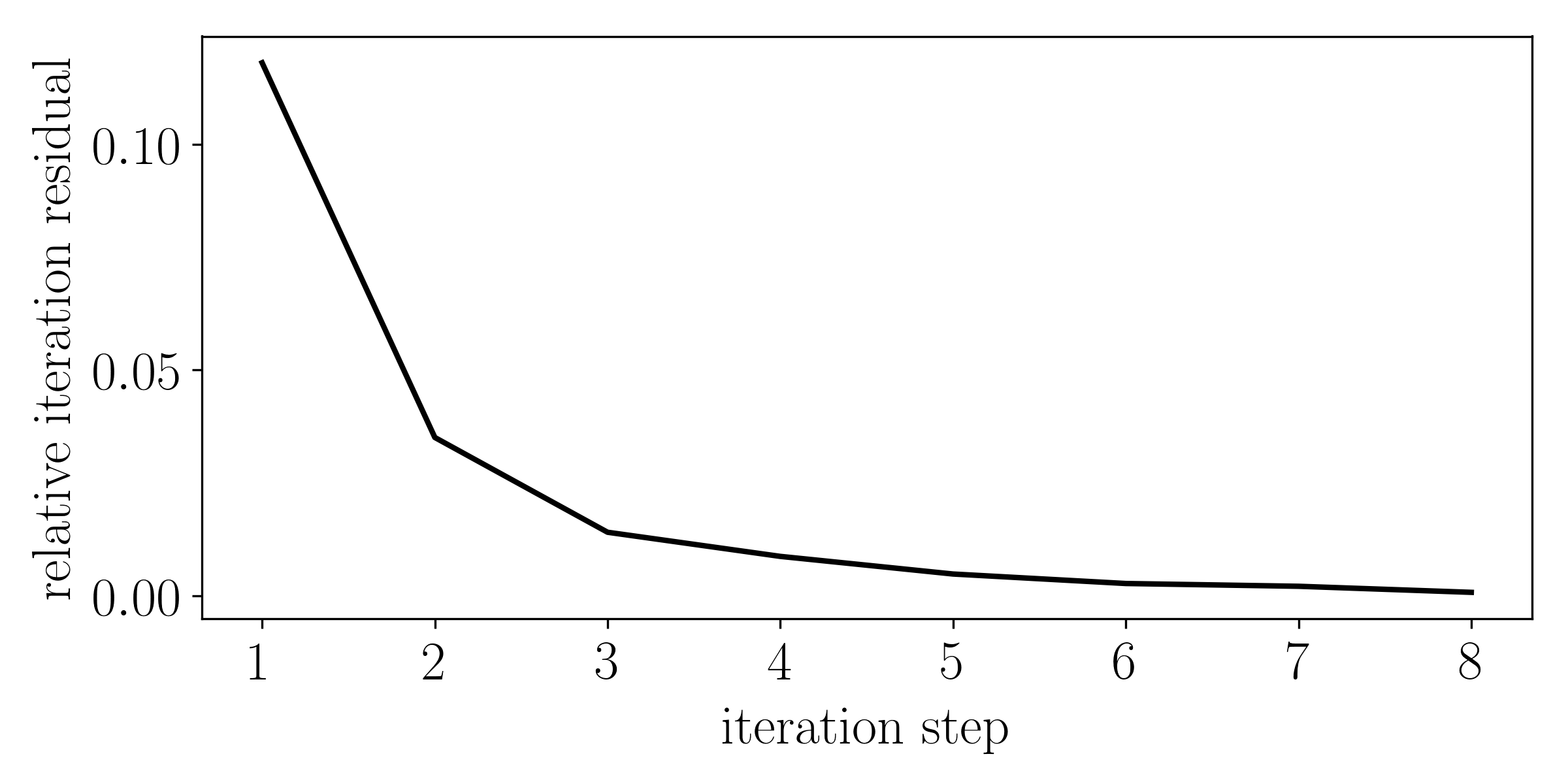}
    		\caption{EnRML}
    \end{subfigure}
    \begin{subfigure}[b]{0.45\linewidth}
    		\includegraphics[width=\textwidth]{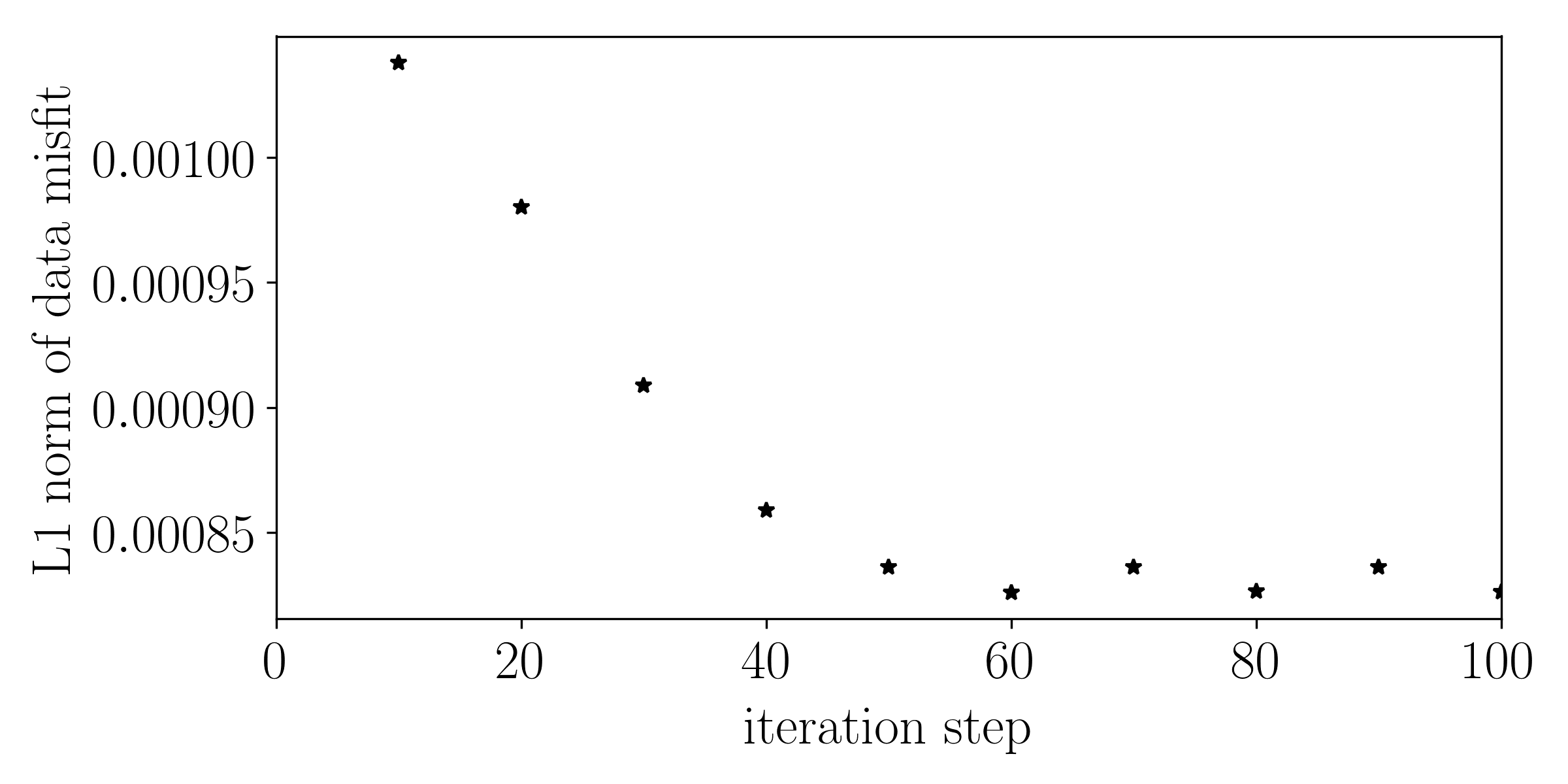}
    		\caption{EnKF-MDA}
    \end{subfigure}
    \caption{Convergence plot of EnRML and EnKF-MDA with $50$ samples}
    \label{fig:converge_plot}
\end{figure}

We also compare the three data assimilation methods in convergence speed. 
The convergence criteria for the three methods are different. Concretely, EnKF and the EnRML method are considered to be converged when the iteration residual in data misfit between the two adjacent iterations is less than $1\times 10^{-3}$, while EnKF-MDA has to reach the predefined maximum iteration number $N_\text{mda}$. From our numerical tests, the EnKF does not converge and stops at the maximum iteration number~$100$. 
Fig.~\ref{fig:converge_plot} presents the evolution of iteration residual for the EnRML method and the convergence plot of the maximum iteration number~$N_\text{mda}$ for EnKF-MDA with $50$ samples.
It can be seen that the EnRML method converges in~$8$ iterations, while EnKF-MDA need at least~$50$ iterations to converge in the maximum iteration number~$N_\text{mda}$, which suggests that EnRML outperforms the EnKF-MDA in convergence speed.

Further, we conduct the data assimilation with $200$, $800$, and $3200$ samples to investigate the effects of sample size.
We use relative data misfit between posterior mean
$\mathsf{H\overline{X}}_i$ and truth~$U^\text{DNS}$ normed by that of prior, to evaluate the posterior results. It can be formulated as
\begin{equation}
    \frac{\parallel \mathcal{H}[\overline{X}]_i - U^\text{DNS} \parallel_\text{L2}}{\parallel \mathcal{H}[\overline{X}]_0 - U^\text{DNS} \parallel_\text{L2}} \text{.}
\end{equation}
Also, the relative standard deviation of the posterior samples is computed in a similar manner to evaluate the reduction of uncertainty after assimilating observation data.
The results with different samples are summarized in Table~\ref{tab:summary_diff_sample}.
It can be seen that EnKF can achieve the best data fit among the three methods but underestimates the variance of the posterior samples.
EnRML and EnKF-MDA not only can improve the data misfit but also provide the statistical information. Comparing between the EnRML and EnKF-MDA, EnRML can provide better data match and preserve larger variance of the posterior.
With large sample size, the posterior variance will be increased for all the three methods since more samples can cover more statistical information.
However, the data misfit will be inferior to that with the small sample size.
That is likely due to the capping error. 
When we perform the Bayesian update, some samples may lead to the updated $(\eta,\xi)$ out of the square $[-1, 1] \times [-1, 1]$ shown in Fig.~\ref{fig:natural_coor}. To ensure physical reliability, we bound any sample outside the square by fixing them at the edges.
With large samples, more samples may jump out of the physical range and need to be capped, which likely causes large errors between the posterior mean and data. Better methods to ensure the physical realizability need to be investigated but are outside the scope of this work.
      \begin{table}[!htb]
        \caption{Summary of the relative data misfit and the relative standard deviation of posterior samples with different ensemble sizes.}
        \label{tab:summary_diff_sample}
        \begin{tabular}{l | l l l}
        \hline
        \hline
        sample size $N=50$ & EnKF & EnRML & EnKF-MDA \\
        \hline
        relative data misfit & $15.2\%$ & $23.2\%$ & $30.0\%$\\
        relative std of ensemble & $16.1\%$ & $36.9\%$ & $30.5\%$\\
        \hline
        sample size $N=200$ &  &  & \\
        \hline
        relative data misfit & $21.7\%$ & $29.2\%$ & $36.6\%$\\
        relative std of ensemble & $17.5\%$ & $40.9\%$ & $31.8\%$\\
        \hline
        sample size $N=800$ & & & \\
        \hline
        relative data misfit & $33.7\%$ & $35.9\%$ & $46.1\%$\\
        relative std of ensemble & $22.5\%$ & $44.4\%$ & $36.2\%$\\
        \hline
        sample size $N=3200$ & & & \\
        \hline
        relative data misfit & $37.9\%$ & $38.2\%$ & $49.3\%$\\
        relative std of ensemble & $23.2\%$ & $49.2\%$ & $38.8\%$\\
        \hline
        \end{tabular}
      \end{table}

\section{Conclusion}
\label{sec:conclusion}
This paper evaluates the performance of three widely used iterative ensemble methods (EnKF, EnRML, and EnKF-MDA) for UQ problems in steady cases.
We summarize the derivations of these ensemble methods from an optimization viewpoint.
The iterative EnKF method performs several full Gauss--Newton steps during which same data is repeatedly used for the stationary scenario. 
The EnRML method and EnKF-MDA can iteratively approach to the optimal point with Gauss--Newton method or likelihood recursion, avoiding the data overuse and alleviating the effects of linearization approximation simultaneously.
From the numerical investigation for a scalar case, we investigate the effects of small ensemble sizes.
The results show that the EnRML method and EnKF-MDA can provide a satisfactory estimation on the posterior uncertainty with small ensemble size but remain inferior to that with large ensemble size. 
This is because the small ensemble size is not sufficient to describe the statistical information and increases the error in the estimation of the model gradient.
This deficiency may be alleviated by using the localization technique, and will be further investigated in future work.
The comparison results for both the scalar case and CFD case show that the posterior mean with all the three methods can have a good agreement with benchmark data.
However, the iterative form of EnKF discussed here which uses the same data repeatedly for steady problems can prompt the data fit but underestimate the posterior uncertainty.
The other two methods, EnRML and EnKF-MDA, are capable of giving an estimation of posterior uncertainty. Based on our comparison study, the EnRML method is recommended since it can converge fast and provide the statistical information even in complicated CFD cases. The applicability of these ensemble methods for unsteady CFD applications will be investigated in future studies.

\section*{Acknowledgements}
The authors would like to thank the reviewers for their constructive and valuable comments, which helped us to improve the quality and clarity of this manuscript.

\appendix
\section{Derivation of EnKF}
\label{appendixA}
The cost function and its gradient for the iterative EnKF are formulated as
\begin{subequations}
\begin{align}
    J &= \frac{1}{2}\left (\mathsf{x}_{n,j}^\text{a}-\mathsf{x}_{n,j}^\text{f} \right)^\top \mathsf{P}_n^{-1} \left (\mathsf{x}_{n,j}^\text{a}-\mathsf{x}_{n,j}^\text{f}\right)+\frac{1}{2}\left(\mathcal{H}[\mathsf{x}_{n,j}^\text{a}]-\mathsf{y}_j\right)^\top \mathsf{R}^{-1}\left(\mathcal{H}[\mathsf{x}_{n,j}^\text{a}]-\mathsf{y}_j\right), \\
    \label{eq:grad_EnKF_A}
    \frac{\partial J}{\partial \mathsf{x}_{n,j}^\text{a}} &= \mathsf{P}_n^{-1}\left(\mathsf{x}_{n,j}^\text{a}-\mathsf{x}_{n,j}^\text{f}\right)+\mathcal{H}'[\mathsf{x}_{n,j}^\text{a}]^\top \mathsf{R}^{-1}\left(\mathcal{H} [\mathsf{x}_{n,j}^\text{a}]-\mathsf{y}_j\right).
\end{align}
\end{subequations}
We approximate the unknown terms $\mathcal{H}[\mathsf{x}^\text{a}]$ and $\mathcal{H'}[\mathsf{x}^\text{a}]$ in Eq.~\eqref{eq:grad_EnKF_A} with the linear assumption as
\begin{subequations}
\begin{align}
    \label{eq:linear_assumption1_A}
    &\mathcal{H}[\mathsf{x}_j^\text{a}] \approx \mathcal{H}[\mathsf{x}_j^\text{f}] + \mathcal{H'}[\mathsf{x}_j^\text{a}]\left(\mathsf{x}_j^\text{a}-\mathsf{x}_j^\text{f}\right),\\
    \label{eq:linear_assumption2_A}
    &\mathcal{H'}[\mathsf{x}_j^\text{a}] \approx \mathcal{H'}[\mathsf{x}_j^\text{f}] + \mathcal{H''}[\mathsf{x}_j^\text{f}]\left(\mathsf{x}_j^\text{a} - \mathsf{x}_j^\text{f}\right),
\end{align}
\label{eq:linear_assumption_A}
\end{subequations}
where the second derivation can be neglected. Further, we set the gradient of the cost function to be zero and
substitute with Eq.~\eqref{eq:linear_assumption_A} as
\begin{equation}
\label{eq:zero_grad1_EnKF_A}
    \mathsf{P}_n^{-1}\left(\mathsf{x}_{n,j}^\text{a} -\mathsf{x}_{n,j}^\text{f}\right)= - \mathcal{H'}[\mathsf{x}_{n,j}^\text{f}]^\top \mathsf{R}^{-1}\left(\mathcal{H}[\mathsf{x}_{n,j}^\text{f}]+\mathcal{H'}[\mathsf{x}_{n,j}^\text{f}]\left(\mathsf{x}_{n,j}^\text{a} - \mathsf{x}_{n,j}^\text{f}\right)-\mathsf{y}_j\right).
\end{equation}
We expand $\mathcal{H}[\mathsf{x}]$ around the ensemble mean as
\begin{subequations}
\label{eq:ensemble_model_gradient_A}
\begin{align}
	&\mathcal{H}[\mathsf{x}_j^\text{f}] \approx \mathcal{H}[\bar{\mathsf{X}}^\text{f}] + \mathcal{H'}[\mathsf{x}_j^\text{f}] \left(\mathsf{x}_j^\text{f}-\bar{\mathsf{X}}^\text{f}\right).
\end{align}
\end{subequations}
Afterwards, we assume that $\mathcal{H}[\mathsf{x}]= \mathsf{H x}$, where $\mathsf{H}$ is the tangent linear operator. The model function gradient $\mathcal{H'}[\mathsf{x}^\text{f}]$ can be estimated directly with the linear operator $\mathsf{H}$ based on Eq.~\ref{eq:ensemble_model_gradient_A}. Hence, Eq.~\eqref{eq:zero_grad1_EnKF_A} can be formulated and rearranged as
\begin{subequations}
\begin{align}
    & \mathsf{P}_n^{-1} \left(\mathsf{x}_{n,j}^\text{a} -\mathsf{x}_{n,j}^\text{f}\right)= - \mathsf{H}^\top \mathsf{R}^{-1}\left(\mathsf{H} \mathsf{x}_{n,j}^\text{f}+\mathsf{H}(\mathsf{x}_{n,j}^\text{a} - \mathsf{x}_{n,j}^\text{f})-\mathsf{y}_j\right), \\
    & \mathsf{x}_{n,j}^\text{a} = \mathsf{x}_{n,j}^\text{f} + \mathsf{P}_n\left(I + \mathsf{H}^\top \mathsf{R}^{-1}\mathsf{HP}_n\right)^{-1}\mathsf{H}^\top \mathsf{R}^{-1}\left(\mathsf{y}_j - \mathsf{H} \mathsf{x}_{n,j}^\text{f}\right).
    \label{eq:zero_grad2_EnKF_A}
\end{align}
\end{subequations}
Set $\mathsf{Q} = \mathsf{R}^{-1}\mathsf{HP}_n$ and we have:
\begin{subequations}
\begin{align}
    &\mathsf{H}^\top \left(I + \mathsf{Q H}^\top \right) = \left(I + \mathsf{H}^\top \mathsf{Q}\right)\mathsf{H}^\top, \\ 
    \label{Identity formula_A}
    &\left(I + \mathsf{H}^\top \mathsf{Q}\right)^{-1}\mathsf{H}^\top = \mathsf{H}^\top \left(I + \mathsf{QH}^\top\right)^{-1}.
\end{align}
\end{subequations}
Now back to Eq.~\eqref{eq:zero_grad2_EnKF_A},  substituting $(I + \mathsf{H}^\top \mathsf{R}^{-1}\mathsf{HP}_n)^{-1}\mathsf{H}^\top$ with $\mathsf{H}^\top(I + \mathsf{R}^{-1}\mathsf{HP}_n\mathsf{H}^\top)^{-1}$ based on Eq.~\eqref{Identity formula_A}, we can derive:
\begin{subequations}
\begin{align}
    & \mathsf{x}_{n,j}^\text{a} = \mathsf{x}_{n,j}^\text{f} + \mathsf{P}_n \mathsf{H}^\top \left(I + \mathsf{R}^{-1}\mathsf{H P}_n \mathsf{H}^\top \right)^{-1}\mathsf{R}^{-1}\left(\mathsf{y}_j - \mathsf{H x}_{n,j}^\text{f}\right),\\
    \label{eq:EnKF_A}
    & \mathsf{x}_{n,j}^\text{a} = \mathsf{x}_{n,j}^\text{f}+ \mathsf{P}_n \mathsf{H}^\top \left( \mathsf{R} + \mathsf{H P}_n \mathsf{H}^\top \right)^{-1}\left(\mathsf{y}_j-\mathsf{H x}_{n,j}^\text{f}\right).
\end{align}
\end{subequations}
Eq.~\eqref{eq:EnKF_A} is the iterative formulation for the analysis step of the EnKF method.

\section{Derivation of EnRML}
\label{appendixB}
To derive the analysis scheme of ensemble randomized maximal likelihood method, we start from the gradient and Hessian of the cost function as
\begin{subequations}
\begin{align}
    & \frac{\partial J}{\partial \mathsf{x}_{l,j}}=\mathsf{P}_0^{-1}\left(\mathsf{x}_{l,j}-\mathsf{x}_{0,j}\right)+\mathcal{H}'[\mathsf{x}_{l,j}]^\top \mathsf{R}^{-1}\left(\mathcal{H} [\mathsf{x}_{l,j}]-\mathsf{y}_j\right),\\
    & \frac{\partial^2 J}{\partial^2 \mathsf{x}_{l,j}} = \mathsf{P}_0^{-1}+ \mathcal{H}' [\mathsf{x}_{l,j}]^\top \mathsf{R}^{-1} \mathcal{H}' [\mathsf{x}_{l,j}].
\end{align}
\end{subequations}
In the EnRML method, the state vector $\mathsf{x}$ is updated with the Gauss--Newton method as
\begin{equation}
    \label{eq:Gauss_Newton_appendix}
    \mathsf{x}_{l,j}^\text{a} = \mathsf{x}_{l,j}^\text{f} - \gamma\left(\frac{\partial^2 J}{\partial^2 \mathsf{x}_{l,j}^\text{f}}\right)^{-1}{\frac{\partial J}{\partial \mathsf{x}_{l,j}^\text{f}}}.
\end{equation}
Through directly introducing the gradient and Hessian formulation into Eq.~\eqref{eq:Gauss_Newton_appendix}, we can have
\begin{equation}
\begin{aligned}
    \mathsf{x}_{l,j}^\text{a} = \mathsf{x}_{l,j}^\text{f} & - \gamma \left(\mathsf{P}_0^{-1}+\mathcal{H'} [\mathsf{x}_{l,j}^\text{f}]^\top \mathsf{R}^{-1} \mathcal{H'}[ \mathsf{x}_{l,j}^\text{f}]\right)^{-1}{\left(\mathsf{P}_0^{-1}(\mathsf{x}_{l,j}^\text{f}-\mathsf{x}_{0,j}^\text{f})+\mathcal{H'} [\mathsf{x}_{l,j}^\text{f}]^\top \mathsf{R}^{-1}(\mathcal{H} [\mathsf{x}_{l,j}^\text{f}] - \mathsf{y}_j)\right)},\\
    \label{eq:EnRMLupdate1_A}
    & - \gamma \left(I+\mathsf{P}_0\mathcal{H'} [\mathsf{x}_{l,j}^\text{f}]^\top \mathsf{R}^{-1}\mathcal{H'} [\mathsf{x}_{l,j}^\text{f}]\right)^{-1} \left(\mathsf{x}_{l,j}^\text{f}-\mathsf{x}_{0,j}^\text{f} + \mathsf{P}_0 \mathcal{H'}[\mathsf{x}_{l,j}^\text{f}]^\top \mathsf{R}^{-1}\left(\mathcal{H} [\mathsf{x}_{l,j}^\text{f}]-\mathsf{y}_j\right)\right).
\end{aligned}
\end{equation}
By expanding the last term, we obtain
\begin{equation}
\begin{aligned}
    \label{eq:EnRMLupdate2_A}
    \mathsf{x}_{l,j}^\text{a}  = & \mathsf{x}_{l,j}^\text{f} - \gamma \left(I+\mathsf{P}_0 \mathcal{H'} [\mathsf{x}_{l,j}^\text{f}]^\top \mathsf{R}^{-1}\mathcal{H'} [\mathsf{x}_{l,j}^\text{f}]\right)^{-1}\left(\mathsf{x}_{l,j}^\text{f}-\mathsf{x}_{0,j}^\text{f}\right) \\&  - \gamma \left(I+\mathsf{P}_0 \mathcal{H'} [\mathsf{x}_{l,j}^\text{f}]^\top \mathsf{R}^{-1}\mathcal{H'} [\mathsf{x}_{l,j}^\text{f}]\right)^{-1} \mathsf{P}_0\mathcal{H'} [\mathsf{x}_{l,j}^\text{f}]^\top \mathsf{R}^{-1}\left(\mathcal{H}[\mathsf{x}_{l,j}^\text{f}]-\mathsf{y}_j\right).
\end{aligned}
\end{equation}
We can further derive from \eqref{eq:EnRMLupdate2_A} via Woodbury formula as follows:
\begin{equation}
\begin{aligned}
    \label{eq:EnRMLupdate3_A}
    \mathsf{x}_{l,j}^\text{a}  = & \mathsf{x}_{l,j}^\text{f} - \gamma \left(I-\mathsf{P}_0 \mathcal{H'} [\mathsf{x}_{l,j}^\text{f}]^\top \left(\mathsf{R}+\mathcal{H'} [\mathsf{x}_{l,j}^\text{f}]\mathsf{P}_0\mathcal{H'} [\mathsf{x}_{l,j}^\text{f}]^\top \right)^{-1}\mathcal{H'} [\mathsf{x}_{l,j}^\text{f}]\right)\left(\mathsf{x}_{l,j}^\text{f}-\mathsf{x}_{0,j}^\text{f}\right) \\&  - \gamma \left(I+\mathsf{P}_0\mathcal{H'} [\mathsf{x}_{l,j}^\text{f}]^\top \mathsf{R}^{-1}\mathcal{H'} [\mathsf{x}_{l,j}^\text{f}]\right)^{-1} \mathsf{P}_0\mathcal{H'} [\mathsf{x}_{l,j}^\text{f}]^\top \mathsf{R}^{-1}\left(\mathcal{H}[\mathsf{x}_{l,j}^\text{f}]-\mathsf{y}_j\right).
\end{aligned}
\end{equation}
After expanding the second term at right hand and rearranging, we can have
\begin{equation}
\begin{aligned}
    \label{eq:EnRMLupdate4_A}
     \mathsf{x}_{l,j}^\text{a} = &\gamma \mathsf{x}_{0,j}^\text{f} +\left(1-\gamma\right) \mathsf{x}_{l,j}^\text{f}+\gamma \mathsf{P}_0 \mathcal{H}'[\mathsf{x}_{l,j}^\text{f}]^\top \left(\mathsf{R}+\mathcal{H}' [\mathsf{x}_{l,j}^\text{f}]\mathsf{P}_0\mathcal{H}' [\mathsf{x}_{l,j}^\text{f}]^\top\right)^{-1}\mathcal{H}' [\mathsf{x}_{l,j}^\text{f}]\left(\mathsf{x}_{l,j}^\text{f}-\mathsf{x}_{0,j}^\text{f}\right) \\
     & - \gamma \left(I+\mathsf{P}_0\mathcal{H'} [\mathsf{x}_{l,j}^\text{f}]^\top \mathsf{R}^{-1}\mathcal{H'} [\mathsf{x}_{l,j}^\text{f}]\right)^{-1} \mathsf{P}_0 \mathcal{H'} [\mathsf{x}_{l,j}^\text{f}]^\top \mathsf{R}^{-1}\left(\mathcal{H}[\mathsf{x}_{l,j}^\text{f}]-\mathsf{y}_j\right)
\end{aligned}
\end{equation}
Set $\mathsf{Q}=\mathsf{P}_0\mathcal{H'}[\mathsf{x}]^\top$, and we deduce
\begin{subequations}
\begin{align}
    & \mathsf{QR}^{-1}\left(\mathsf{R} + \mathcal{H'}[\mathsf{x}]\mathsf{Q}\right) =\left(I+\mathsf{QR}^{-1}\mathcal{H'}[\mathsf{x}]\right)\mathsf{Q},\\  
    \label{eq:Identity_formula_EnRML_A}
    & \left(I+\mathsf{QR}^{-1}\mathcal{H'}[\mathsf{x}]\right)^{-1}\mathsf{QR}^{-1} = \mathsf{Q}\left(\mathsf{R}+ \mathcal{H'}[\mathsf{x}]\mathsf{Q}\right)^{-1}.
\end{align}
\end{subequations}
Finally, by substituting Eq.~\eqref{eq:Identity_formula_EnRML_A} into Eq.~\eqref{eq:EnRMLupdate4_A} we can obtain the analysis step for the EnRML method as
\begin{equation}
\begin{aligned}
    \label{eq:EnRMLupdate5_A}
    \mathsf{x}_{l,j}^\text{a} = \gamma \mathsf{x}_{0,j}^\text{f} + \left(1- \gamma\right) \mathsf{x}_{l,j}^\text{f}-\gamma \mathsf{P}_0\mathcal{H}' [\mathsf{x}_{l,j}^\text{f}]^\top\left(\mathsf{R}+\mathcal{H}'[\mathsf{x}_{l,j}^\text{f}]^\top \mathsf{P}_0 \mathcal{H}' [\mathsf{x}_{l,j}^\text{f}]\right)^{-1}\left(\mathcal{H} [\mathsf{x}_{l,j}^\text{f}]-\mathsf{y}_j-\mathcal{H}'[\mathsf{x}_{l,j}^\text{f}]\left(\mathsf{x}_{l,j}^\text{f}-\mathsf{x}_{0,j}^\text{f}\right)\right).
\end{aligned}
\end{equation}

\end{document}